\documentclass{article}

 \usepackage[preprint]{neurips_2025}
\usepackage{tikz}
\usetikzlibrary{shapes.geometric, arrows.meta, positioning, calc, fit, backgrounds}

\usepackage[utf8]{inputenc} 
\usepackage[T1]{fontenc}    
\usepackage{hyperref}       
\usepackage{url}            
\usepackage{amsfonts}       
\usepackage{nicefrac}       
\usepackage{microtype}      
\setcitestyle{numbers,square}

\usepackage[table]{xcolor}
\usepackage{booktabs, multirow, colortbl, array, makecell}
\definecolor{bestrow}{HTML}{E8F5E9}
\definecolor{headerblue}{HTML}{E3F2FD}
\definecolor{lightgray}{HTML}{F5F5F5}

\usepackage{comment}
\usepackage{tcolorbox}
\tcbuselibrary{breakable,skins}
\tcbuselibrary{listings}
\usepackage{listings}

\usepackage{algorithm}
\usepackage{algorithmic}

\usepackage{xcolor}

\newtcolorbox{llmexample}[2][]{
  breakable,
  colback=gray!3,
  colframe=gray!50,
  fonttitle=\bfseries,
  title={#2},
  boxrule=0.5pt,
  arc=2pt,
  left=6pt,
  right=6pt,
  top=6pt,
  bottom=6pt,
  #1
}

\lstdefinestyle{jsonstyle}{
  basicstyle=\ttfamily\small,
  breaklines=true,
  columns=fullflexible,
  keepspaces=true,
  showstringspaces=false,
  frame=single,
  framerule=0.3pt
}

\title{EigenData: A Self-Evolving Multi-Agent Platform for Function-Calling Data Synthesis, Auditing, and Repair}

\author{%
  Jiaao Chen \quad
  Jingyuan Qi \quad
  Mingye Gao \quad
  Wei-Chen Wang \quad
  Hanrui Wang \quad
  Di Jin\thanks{Correspondence to Di Jin (\texttt{di@eigenai.com}).}  \\[6pt]
  \textbf{Eigen AI} \\
}

\makeatletter
\renewcommand{\@notice}{}
\makeatother

\begin{document}

\maketitle

\begin{abstract}
Function-calling agents---large language models that invoke tools and APIs---require high-quality, domain-specific training data spanning executable environments, backing databases, and diverse multi-turn trajectories. We introduce \textbf{EigenData}, an integrated, self-evolving platform that automates the full data lifecycle through a multi-agent architecture. A top-level orchestrator, \textbf{EigenCore}, coordinates three specialized sub-systems: \textbf{DatabaseAgent} for realistic domain database construction, \textbf{CodingAgent} for verified executable environment generation with iterative test-debug loops, and \textbf{DataAgent} for multi-turn trajectory synthesis with self-evolving prompt optimization. Cross-component feedback ensures consistency across all artifacts. We apply EigenData to audit and repair the Berkeley Function-Calling Leaderboard (BFCL-V3), identifying systematic errors in function schemas, implementations, and reference trajectories, automatically correcting them through coordinated schema refinement, code-level bug fixes, and trajectory modification, and introducing an outcome-aware evaluation protocol that assesses task success via database-state correctness rather than turn-level trajectory matching. We demonstrate that the repaired benchmark, coupled with outcome-aware metrics, produces model rankings substantially better correlated with human judgments of functional correctness.
\end{abstract}

\section{Introduction}
Function-calling agents—LLMs that can invoke tools and APIs—are rapidly becoming a universal abstraction to automate workflows across domains such as travel booking, customer support, logistics, and enterprise operations~\citep{barres2025tau2benchevaluatingconversationalagents,li2025tooldecathlonbenchmarkinglanguage,luo2025mcpuniversebenchmarkinglargelanguage,wu2025mcpmarkbenchmarkstresstestingrealistic,tbench_2025,drouin2024workarenacapablewebagents,zhou2024webarenarealisticwebenvironment,ye2025maslabunifiedcomprehensivecodebase,tang2025eigen}. 
In realistic deployments, achieving high performance requires in-domain post-training on representative trajectories—whether through supervised fine-tuning (SFT)~\citep{kimiteam2025kimik2openagentic,li2025simulatingenvironmentsreasoningmodels,prabhakar2025apigenmtagenticpipelinemultiturn,liu2024apigenautomatedpipelinegenerating} or reinforcement learning (RL) with outcome-based rewards~\citep{zhuang2025workforceagentr1incentivizingreasoningcapability,deepseekai2025deepseekv32pushingfrontieropen,li2025simulatingenvironmentsreasoningmodels,su2025toolorchestraelevatingintelligenceefficient,lu2025scaling}.
In these settings, data and evaluation are the primary bottlenecks~\citep{yehudai2025surveyevaluationllmbasedagents}: models cannot be meaningfully improved beyond the quality, coverage, and faithfulness of the datasets and metrics used to train and assess them.

Today, most high-quality function-calling agent data still depend on human annotation~\citep{barres2025tau2benchevaluatingconversationalagents,luo2025mcpuniversebenchmarkinglargelanguage}. Annotators must understand domain-specific schemas, reason about multi-step tool use, and maintain global consistency across complex tasks. This process is slow and expensive; as we push for more realistic and diverse scenarios, we face a sharp tradeoff between quality and scale under limited time and budget. Meanwhile, real-world interaction logs are noisy and incomplete, making them difficult to convert into clean training signal. As a result, even widely used benchmarks and training corpora often contain schema inconsistencies, ambiguous user intents, and labels that do not faithfully capture task success~\citep{anthropic_claude_opus_45_system_card_2025,futurehouse2025hle,openai2025swebenchverified}.

To mitigate annotation costs, several works have turned to rule-based or scripted synthetic data generation, including function-calling benchmarks such as BFCL~\citep{patilberkeley,patil2023gorilla} and TOUCAN~\citep{xu2025toucan}, as well as automated pipelines like ToolACE~\citep{toolace2024}. These pipelines can produce large amounts of data, but they are brittle. In practice, they often exhibit (i) schema bugs, where function definitions or implementations are incorrect or inconsistent; (ii) ambiguous or underspecified user intents, where multiple outcomes are plausible but only one is labeled ; and (iii) weak evaluation metrics, typically based on turn-level function-call matching rather than on the actual consequences of agent actions. Such issues can both mislead model selection and obscure genuine progress: an agent may achieve high function-match scores while failing to produce the correct database state or to satisfy user goals.

We argue that closing this gap requires not merely better data generators, but an \emph{end-to-end infrastructure} that can synthesize entire training environments---including executable tool implementations, backing databases, and diverse multi-turn trajectories---while also auditing, repairing, and iteratively refining existing datasets and evaluation protocols. To this end, we introduce \textbf{EigenData}, an integrated, self-evolving platform for function-calling data that spans the full lifecycle from environment construction to trajectory generation and quality assurance.

At its core, EigenData is coordinated by \textbf{EigenCore}, a top-level orchestration module that receives human requests---ranging from ``generate training data for a hotel-booking agent'' to ``audit the BFCL benchmark''---and decomposes them into sub-tasks dispatched to three specialized sub-systems:
\begin{itemize}
    \item \textbf{DatabaseAgent} constructs the domain databases that underpin tool-use environments (Section~\ref{sec:database-agent}). Given a domain specification (e.g., schema descriptions, sample constraints, or natural-language requirements), it generates realistic, internally consistent database instances---including relational tables, JSON stores, or key-value structures---that serve as the ground-truth state against which agent trajectories are evaluated.

    \item \textbf{CodingAgent} produces the executable code layer: environment implementations (e.g., Python simulation environments, REST APIs, or MCP servers) that expose tool interfaces backed by the generated databases, as well as auxiliary scripts such as format-conversion utilities and evaluation harnesses (Section~\ref{sec:coding-agent}). CodingAgent employs a multi-agent, iterative test-debug workflow---generating code, synthesizing unit and integration tests, and invoking a judge-based refinement loop---to ensure that every produced artifact is functionally correct before it enters the data pipeline.

    \item \textbf{DataAgent} is the multi-agent data-generation engine described in detail in Section~\ref{sec:dataagent}. Operating on the environments and databases produced by the preceding sub-systems, it orchestrates a hierarchy of specialized worker agents to synthesize, evaluate, and iteratively refine high-quality multi-turn function-calling trajectories for both SFT and RL.
\end{itemize}
A complete generation workflow from scratch proceeds as follows: the DatabaseAgent first constructs the domain database; the CodingAgent then generates the executable environment that exposes tool APIs backed by that database; and finally, the DataAgent uses both artifacts to produce diverse, verified training trajectories. EigenCore manages this pipeline end to end, handling dependency resolution, cross-component consistency checks, and iterative feedback among sub-systems. Importantly, each sub-system can also be invoked independently---for instance, the DataAgent alone can audit and repair an existing benchmark that already provides environments and databases, which is the focus of the case study in this paper.

In this paper, we present the full EigenData architecture and instantiate it on the BFCL-V3 function-calling benchmark, with a focus on benchmark auditing, repair, and extension that demonstrates the DataAgent sub-system in coordination with the CodingAgent and DatabaseAgent.\footnote{Corrected BFCL-V3 is stored at: https://github.com/eigen-ai-labs/eigendata-corrected-bfcl-v3} Specifically, we use EigenData to (1)~systematically audit and categorize errors in function schemas, implementations, reference trajectories, and user intents; and (2)~repair these errors and disambiguate intents. In parallel, we introduce an outcome-aware evaluation protocol---inspired by Tau-style benchmarks~\citep{barres2025tau2benchevaluatingconversationalagents}---that goes beyond turn-level function matching to explicitly assess whether (i)~the predicted database changes match ground truth, (ii)~key functions have been correctly invoked, and (iii)~critical information has been correctly handled and conveyed. We show that this repaired and extended benchmark, coupled with outcome-based metrics, reveals qualitatively different failure modes and leads to substantially different model rankings than the original BFCL-V3 evaluation.

To make EigenData accessible to practitioners, we have released a command-line interface (CLI) that exposes the platform's core capabilities---including data generation, schema refinement, auditing, and repair---through a unified, scriptable workflow. The CLI and its documentation are available at: \url{https://docs.eigenai.com/products/eigendata-cli/intro}.





\section{Related Work}
\label{sec:related-work}

\paragraph{Function-Calling Benchmarks and Synthetic Data.}
The rapid growth of tool-augmented LLMs has spurred a wave of benchmarks: BFCL~\citep{patilberkeley,patil2023gorilla} evaluates single- and multi-turn function calling, AgentBench~\citep{liu2023agentbench} provides eight agent evaluation environments, and domain-specific suites such as WebArena~\citep{zhou2024webarenarealisticwebenvironment}, SWE-bench~\citep{jimenez2024swebench}, $\tau^2$-Bench~\citep{barres2025tau2benchevaluatingconversationalagents}, and Tool Decathlon~\citep{li2025tooldecathlonbenchmarkinglanguage} target web navigation, software engineering, conversational agents, and long-horizon tasks, respectively. To address the cost of human annotation, several pipelines automate function-calling data production: APIGen~\citep{liu2024apigenautomatedpipelinegenerating} and APIGen-MT~\citep{prabhakar2025apigenmtagenticpipelinemultiturn} generate verifiable datasets through simulated agent--human interplay; ToolACE~\citep{toolace2024} uses tool self-evolution to synthesize diverse APIs; TOUCAN~\citep{xu2025toucan} scales to 1.5M trajectories from real-world MCP environments; and MAG-V~\citep{sengupta2024magv} applies multi-agent generation with verification. While effective at scale, these pipelines typically operate as fixed, linear workflows without the iterative self-correction, cross-component feedback, or end-to-end environment construction (databases, code, trajectories) that EigenData provides.

\paragraph{Multi-Agent Systems for Code Generation and Evaluation.}
Multi-agent architectures have shown promise for complex software tasks: ChatDev~\citep{qian2024chatdev} models software development as multi-role conversation, AgentCoder~\citep{huang2024agentcoder} combines programmer, test designer, and executor agents with iterative refinement, and MASLab~\citep{ye2025maslabunifiedcomprehensivecodebase} provides a unified codebase for LLM-based multi-agent systems. In parallel, LLM-based evaluation has matured since MT-Bench~\citep{zheng2023judgingllmasajudge} demonstrated high agreement with human preferences, with subsequent work exploring agent-level assessment~\citep{yu2025aisjudgeaisrise} and comprehensive evaluation taxonomies~\citep{yehudai2025surveyevaluationllmbasedagents}. EigenData's CodingAgent extends multi-agent code generation with a judge-based fault-attribution mechanism and two-stage testing (unit then workflow), while its DataAgent introduces hierarchical orchestration with LLM-as-judge quality control and programmatic verification functions that can serve as reward models for reinforcement learning~\citep{cheng2025agentr1,zhuang2025workforceagentr1incentivizingreasoningcapability}.

\paragraph{Prompt Optimization and Self-Evolving Agents.}
Automated prompt refinement has evolved from evolutionary search~\citep{fernando2023promptbreeder} to declarative pipeline compilation~\citep{khattab2023dspy} and hierarchical multi-agent optimization~\citep{liu2025hierarchicalmultiagentworkflowszeroshot}. Recent work on self-evolving agents extends this paradigm to entire agent workflows, where systems iteratively improve not just prompts but also tool selection, coordination strategies, and quality criteria~\citep{li2025simulatingenvironmentsreasoningmodels}. EigenData's two-phase self-evolving process (Section~\ref{sec:eigendata-two-phase}) draws on these ideas: Phase~1 optimizes prompts on a small pilot batch through iterative judge feedback, and Phase~2 scales with continuous quality monitoring, combining the efficiency of pilot-based tuning with the robustness of production-time verification.

\section{EigenData: An Integrated Platform for Function-Calling Data}
\label{sec:eigendata}

\begin{figure}[t]
    \centering
    \includegraphics[width=1.0\linewidth]{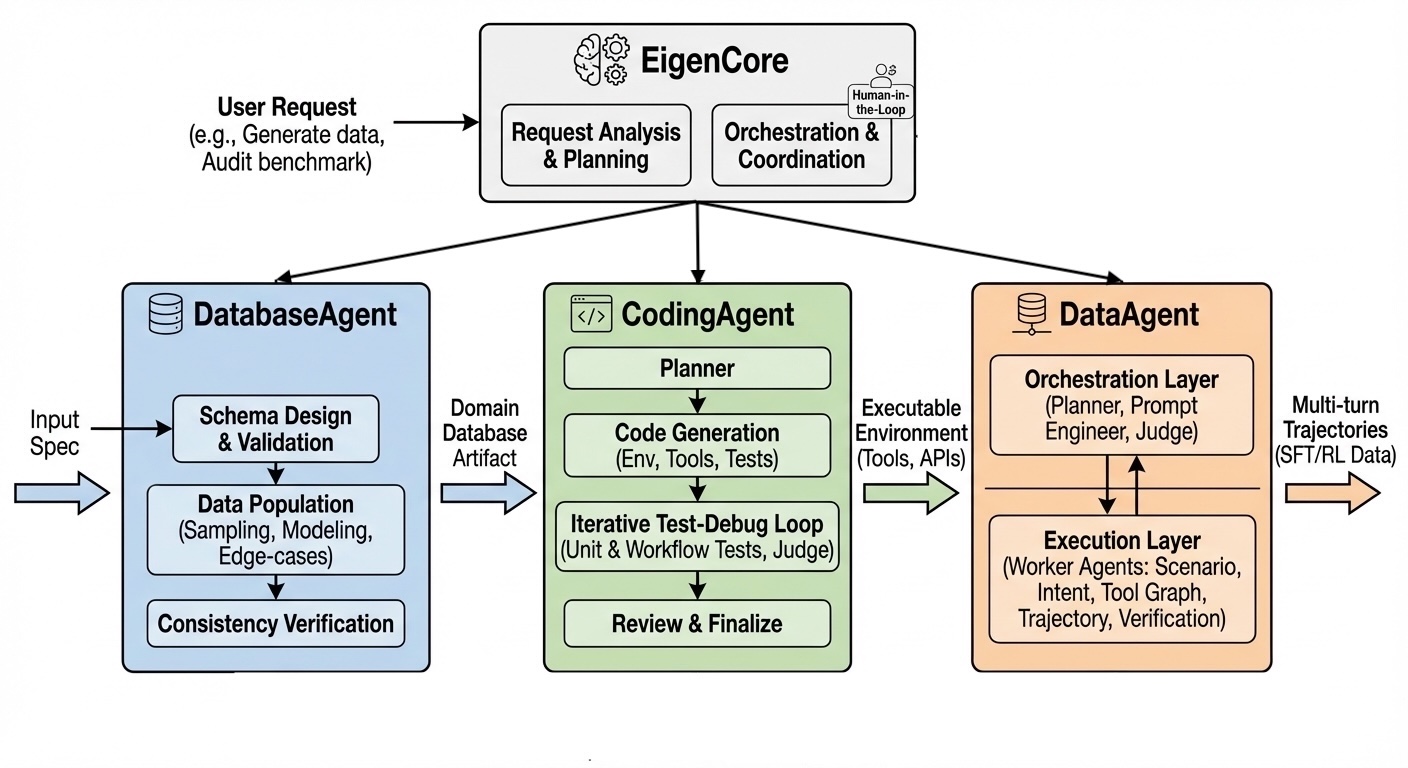}
    \caption{Overview of the EigenData platform. EigenCore receives user requests and orchestrates three specialized sub-systems: DatabaseAgent for domain database construction, CodingAgent for environment and tool generation, and DataAgent for multi-turn trajectory synthesis. Artifacts flow left to right through the canonical pipeline, with cross-component feedback enabling targeted repair without full-pipeline restarts.}
    \label{fig:eigendata-arch}
\end{figure}

We present \textbf{EigenData}, an integrated platform for automated agentic data generation that spans the full lifecycle from environment construction to trajectory synthesis and quality assurance. EigenData is organized around a top-level orchestration module, \textbf{EigenCore} (Section~\ref{sec:eigencore}), which receives human requests and coordinates three specialized sub-systems: \textbf{DatabaseAgent} (Section~\ref{sec:database-agent}) for domain database construction, \textbf{CodingAgent} (Section~\ref{sec:coding-agent}) for executable environment and tool generation, and \textbf{DataAgent} (Section~\ref{sec:dataagent}) for multi-turn trajectory synthesis. Each sub-system is itself a multi-agent system composed of specialized worker agents; EigenCore manages their interplay, enforces cross-component consistency, and supports both end-to-end generation from scratch and targeted operations on existing artifacts.

In contrast to single-LLM prompting or static pipelines~\citep{liu2024apigenautomatedpipelinegenerating,prabhakar2025apigenmtagenticpipelinemultiturn}, our multi-agent architecture---inspired by recent advances in collaborative agent systems~\citep{qian2024chatdev,ye2025maslabunifiedcomprehensivecodebase}---enables dynamic workflow planning, prompt self-evolution, complex task decomposition, and self-correction, yielding realistic, diverse, and domain-specific data that meets rigorous quality criteria. Moreover, the framework is designed to be domain-agnostic: by simply providing domain documentation or APIs as context, the same pipeline can be applied to different domains effortlessly, adapting its behavior to the nuances of each new task.

\subsection{EigenCore: Top-Level Orchestration}
\label{sec:eigencore}

EigenCore is the user-facing entry point and top-level controller of the EigenData platform as illustrated in Figure~\ref{fig:eigendata-arch}. It accepts natural-language requests that may range from narrow tasks (``repair the schemas in BFCL category~X'') to broad directives (``generate a complete training environment and 1{,}000 trajectories for an airline-booking agent''), and decomposes each request into a dependency-aware execution plan over the three sub-systems.

\paragraph{Request Understanding and Task Decomposition.}
Upon receiving a user request, EigenCore first invokes a \emph{Request Analysis} module that classifies the request type (e.g., full-pipeline generation, environment-only construction, data auditing, benchmark extension) and extracts key parameters such as target domain, scale, quality constraints, and any user-supplied artifacts (existing databases, schemas, or code). Based on this analysis, EigenCore produces a \emph{task graph}---a directed acyclic graph of sub-tasks with explicit data dependencies. For example, a from-scratch generation request yields the canonical three-stage pipeline: DatabaseAgent $\rightarrow$ CodingAgent $\rightarrow$ DataAgent, whereas a benchmark-repair request may invoke only the DataAgent on existing artifacts.

\paragraph{Cross-Component Coordination.}
EigenCore manages the flow of artifacts between sub-systems, ensuring that outputs from one stage (e.g., the database schema produced by DatabaseAgent) are validated and correctly consumed by the next (e.g., CodingAgent's environment generator). When downstream stages detect inconsistencies---such as a generated environment referencing database fields that do not exist---EigenCore routes structured feedback to the responsible upstream sub-system and triggers targeted re-generation, avoiding full-pipeline restarts.

\paragraph{Human-in-the-Loop Interface.}
EigenCore supports interactive operation, allowing users to inspect intermediate artifacts, approve or reject sub-system outputs, adjust parameters, and steer the generation process. This is particularly valuable during the initial setup of a new domain, where user oversight can rapidly calibrate the system's behavior. For large-scale batch generation, EigenCore operates autonomously once the user approves the initial configuration.

\subsubsection{Three-Agent Coordination}
\label{sec:three-agent-coordination}

As illustrated in Figure~\ref{fig:eigendata-arch}, EigenCore coordinates three specialized sub-systems---DatabaseAgent, CodingAgent, and DataAgent---each responsible for a distinct layer of the training-environment stack. Their interaction follows a dependency-driven artifact pipeline, and EigenCore can selectively invoke any subset depending on the user's request. To assemble a complete task specification, EigenCore employs a multi-turn conversational interface that progressively elicits the required command parameters (e.g., target domain name, MCP server URL, and final sample count) across successive turns. Three agents implement this interface: an \textbf{ExtractionAgent} and a \textbf{ConversationAgent} drive the interactive collection loop, while a \textbf{RequestBuilder} synthesizes the final directive once all parameters have been supplied.

Each turn begins with the \textit{ExtractionAgent}, which receives the user's message, the conversation history, the parameters accumulated thus far, and the schema of available commands. The agent determines whether the message specifies a concrete command: if so, it identifies the command, extracts any supplied arguments (e.g., domain name, sample count), and assigns the turn status \textit{collecting}; if the message pertains to the system but the intended command remains ambiguous, the status is \textit{no-task}; if the message falls outside system functionality (e.g., a general knowledge question), the status is \textit{chitchat}. The \textit{ConversationAgent} then generates a response conditioned on this status---serving as a general-purpose assistant in chitchat state, prompting the user to clarify their intent in no-task state, or soliciting the specific missing parameters in collecting state (e.g., ``How many samples would you like to generate?''). This two-agent loop iterates until all required parameters have been supplied, at which point the status transitions to \textit{ready}.

Once all required parameters are present, EigenCore invokes the \textit{RequestBuilder}, which consumes the full set of accumulated parameters and the conversation history to produce a single, well-structured natural-language instruction that unambiguously captures the user's intent (e.g., ``Generate 200 multi-turn airline booking dialogs with function calls, using the provided MCP server schema''). The \textit{ConversationAgent} presents this synthesized request to the user for final confirmation. Once confirmed, EigenCore exits the conversational loop and routes the completed specification to the appropriate sub-system: DatabaseAgent for domain-database construction, CodingAgent for environment generation, DataAgent for trajectory synthesis, or any combination thereof as determined by the task graph produced during request analysis.

\subsection{DatabaseAgent: Domain Database Construction}
\label{sec:database-agent}

\begin{figure}[t]
    \centering
    \includegraphics[width=1.0\linewidth]{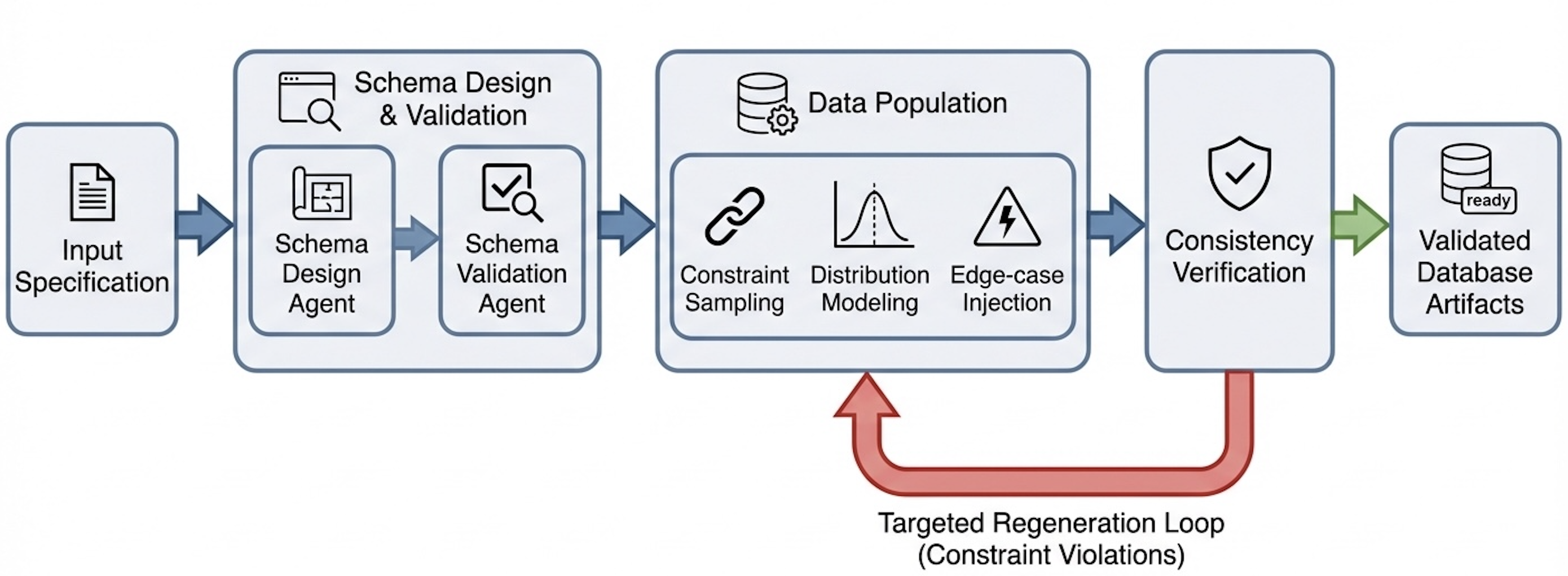}
    \caption{Internal architecture of the DatabaseAgent. The pipeline proceeds from input specification through schema design, validation, data population (with constraint-aware sampling, distribution modeling, and edge-case injection), and consistency verification. A targeted regeneration loop handles any constraint violations detected during verification.}
    \label{fig:database-agent}
\end{figure}

The DatabaseAgent is responsible for constructing the domain databases that serve as the ground-truth state for tool-use environments as illustrated by Figure~\ref{fig:database-agent}. High-quality databases are critical: they determine the space of possible user queries, constrain valid tool-call sequences, and provide the reference state against which agent trajectories are evaluated. Manually creating realistic databases for each new domain is labor-intensive and error-prone; the DatabaseAgent automates this process.

\subsubsection{Input Specification}
The DatabaseAgent accepts domain specifications in multiple forms, from high-level natural-language descriptions (e.g., ``an airline database with flights, passengers, bookings, and seat assignments'') to detailed schema definitions with constraints, cardinalities, and example records. When the specification is underspecified, the DatabaseAgent infers reasonable defaults by leveraging the LLM's world knowledge about the target domain, then presents the inferred schema to the user (via EigenCore) for confirmation before proceeding to data population.

\subsubsection{Schema Design and Validation}
A \textbf{SchemaDesignAgent} translates the input specification into a formal database schema---defining tables (or collections), columns, data types, primary and foreign keys, uniqueness constraints, and value ranges. The agent reasons about domain semantics to ensure referential integrity (e.g., every booking references a valid flight and passenger) and realistic data distributions (e.g., flight prices vary by route, class, and season). The proposed schema is validated by a \textbf{SchemaValidationAgent} that checks for structural soundness, constraint satisfiability, and coverage of the tool APIs that will be built on top of it.

\subsubsection{Data Population}
A \textbf{DataPopulationAgent} generates synthetic records that conform to the validated schema. Rather than producing uniformly random data, it employs domain-aware generation strategies:
\begin{itemize}
    \item \textbf{Constraint-aware sampling}: Generates records that satisfy all inter-table constraints (e.g., departure times precede arrival times; seat assignments do not exceed aircraft capacity).
    \item \textbf{Distribution modeling}: Produces realistic value distributions informed by domain priors (e.g., most hotel bookings are 1--3 nights; flight prices follow route-dependent distributions).
    \item \textbf{Edge-case injection}: Deliberately includes boundary conditions and rare-but-valid scenarios (e.g., sold-out flights, canceled bookings, waitlisted passengers) to ensure that downstream training data covers failure modes.
\end{itemize}

\subsubsection{Consistency Verification}
After population, a \textbf{ConsistencyVerificationAgent} performs end-to-end validation: executing referential-integrity checks, verifying that all constraints hold, and running a suite of representative queries to confirm that the database supports the intended tool-use scenarios. Any violations trigger targeted regeneration of the offending records rather than full repopulation. The final output is a self-contained database artifact (e.g., a JSON file, SQLite database, or structured collection) along with its schema documentation, ready for consumption by the CodingAgent.

\subsection{CodingAgent: Environment and Tool Generation}
\label{sec:coding-agent}

\begin{algorithm}[t]
\caption{CodingAgent: Two-Stage Iterative Test-Debug Pipeline}
\label{fig:coding-agent}
\begin{algorithmic}[1]
\REQUIRE Domain specification $\mathcal{S}$, database $\mathcal{D}$, max iterations $M$, max cycles $N$
\ENSURE Verified environment module, test suites, (optional) MCP server
\medskip
\STATE \textbf{Planner} analyzes $\mathcal{S}$ and $\mathcal{D}$ $\rightarrow$ execution plan with function list and dependency graph
\STATE \textbf{FileGenerationAgent} generates module skeleton and class structure
\medskip
\STATE \textbf{--- Stage 1: Unit Testing (per function) ---}
\FOR{each function $f$ in execution plan}
    \STATE $\mathcal{T}_f \leftarrow$ \textbf{TestingAgent}.generate\_unit\_tests($f$) \COMMENT{happy path, errors, edge cases, types}
    \STATE $\hat{f} \leftarrow$ \textbf{CodingAgent}.generate($f$, $\mathcal{S}$, $\mathcal{D}$)
    \FOR{cycle $c = 1$ \TO $N$}
        \FOR{iteration $i = 1$ \TO $M$}
            \STATE result $\leftarrow$ run\_tests($\hat{f}$, $\mathcal{T}_f$)
            \IF{result = \textsc{Pass}}
                \STATE \textbf{break} both loops \COMMENT{function $f$ passes}
            \ENDIF
            \STATE $\hat{f} \leftarrow$ \textbf{CodingAgent}.fix($\hat{f}$, result.feedback)
        \ENDFOR
        \IF{result $\neq$ \textsc{Pass}}
            \STATE fault $\leftarrow$ \textbf{JudgeAgent}.attribute(result, $\hat{f}$, $\mathcal{T}_f$)
            \IF{fault = \textsc{Tests}}
                \STATE $\mathcal{T}_f \leftarrow$ \textbf{TestingAgent}.regenerate($f$, fault.feedback) \COMMENT{fix faulty tests}
            \ELSE
                \STATE $\hat{f} \leftarrow$ rollback($f$); incorporate judge feedback \COMMENT{retry code from scratch}
            \ENDIF
        \ENDIF
    \ENDFOR
\ENDFOR
\medskip
\STATE \textbf{--- Stage 2: Workflow Testing (dependency groups) ---}
\FOR{each dependency group $G$ in dependency graph}
    \STATE $\mathcal{T}_G \leftarrow$ \textbf{TestingAgent}.generate\_workflow\_tests($G$) \COMMENT{integration tests}
    \REPEAT
        \STATE result $\leftarrow$ run\_tests($G$, $\mathcal{T}_G$)
        \IF{result $\neq$ \textsc{Pass}}
            \STATE identify responsible function(s) $\subseteq G$ or wrong tests; apply targeted fixes
        \ENDIF
    \UNTIL{result = \textsc{Pass} \OR max workflow iterations reached}
\ENDFOR
\medskip
\STATE \textbf{ReviewAgent} performs batch code review over all functions \COMMENT{final quality gate}
\IF{review rejects any function (strict mode)}
    \STATE refine rejected functions and re-validate
\ENDIF
\RETURN environment module, test suites, MCP server (optional), auxiliary scripts
\end{algorithmic}
\end{algorithm}
The CodingAgent produces the executable software layer of the training environment (Algorithm~\ref{fig:coding-agent}): the tool implementations, simulation environments, and auxiliary scripts that expose domain functionality as callable APIs. Given a domain specification and a database (typically produced by the DatabaseAgent), it generates code that is functionally correct, tested, and ready for use by the DataAgent's trajectory-generation pipeline.

\subsubsection{Architecture}
The CodingAgent is itself a multi-agent system organized around an \textbf{EnhancedOrchestrator} that coordinates several specialized agents through an iterative generate-test-debug workflow, extending multi-agent code generation paradigms~\citep{huang2024agentcoder,qian2024chatdev}:

\begin{itemize}
    \item \textbf{Planner}: Analyzes the domain specification and database schema to determine the set of functions to implement, their signatures, dependencies, and the overall module structure. It produces an execution plan that the orchestrator follows.

    \item \textbf{FileGenerationAgent}: Generates complete Python modules from natural-language specifications. For environment files, it produces stateful classes that load the domain database and expose tool functions (e.g., \texttt{search\_flights}, \texttt{book\_hotel}, \texttt{cancel\_reservation}). For MCP (Model Context Protocol) servers, it generates FastMCP-based servers with REST endpoints, session isolation, and auto-expiration. It can also generate format-conversion scripts and evaluation harnesses.

    \item \textbf{CodingAgent (inner)}: The core code-generation agent that implements individual functions. It operates in multiple modes---\emph{generate} (implement from scratch), \emph{correct} (fix bugs), and \emph{refine} (improve working code)---and produces code that follows a consistent error-handling convention: public methods return structured error dictionaries while private methods propagate exceptions.

    \item \textbf{TestingAgent}: Automatically generates unit tests and integration (workflow) tests from function schemas and specifications. Unit tests cover individual function correctness (happy paths, error cases, edge cases, type validation), while workflow tests verify that related functions integrate correctly (e.g., that \texttt{book\_flight} correctly updates the state queried by \texttt{get\_booking}).

    \item \textbf{JudgeAgent}: When tests fail, the JudgeAgent determines whether the fault lies in the generated code or in the generated tests---a critical distinction, since LLM-generated tests can themselves be incorrect. Based on the judge's verdict, either the code or the tests are regenerated with targeted feedback.

    \item \textbf{ReviewAgent}: A quality gate that performs batch code review across all generated functions, checking for consistent patterns, proper state management, and adherence to specifications. In strict mode, rejected functions are automatically refined until they pass review.
\end{itemize}

\subsubsection{Iterative Test-Debug Loop}
The CodingAgent's central mechanism is a two-stage testing pipeline with judge-based refinement (Algorithm~\ref{fig:coding-agent}):

\paragraph{Stage~1: Unit Testing.} For each function, the system (1)~generates unit tests from the function schema, (2)~generates or refines the implementation, (3)~runs the tests, and (4)~if tests fail, enters a judge-based debug loop. This loop has a nested structure: an outer cycle (up to $N$~iterations) governs the unit tests, and an inner cycle (up to $M$~iterations) governs the code implementation. In the inner cycle, the CodingAgent iteratively revises the implementation to pass the current unit tests. If the inner cycle exhausts $M$~attempts without success, the JudgeAgent intervenes to determine fault attribution---deciding whether the failure lies in the code or in the tests themselves. If the tests are deemed faulty, the outer cycle advances: the tests are regenerated with judge feedback, and the inner cycle restarts against the corrected tests. This process continues for up to $N$~outer cycles per function.

\paragraph{Stage~2: Workflow Testing.} After all individual functions pass their unit tests, the system groups functions by their call dependencies and generates integration tests for each group. These tests verify that functions interact correctly---for example, that state mutations from one function are correctly observed by another. Failures trigger targeted fixes to the responsible functions.

\subsubsection{Output Artifacts}
The CodingAgent produces several artifacts:
\begin{itemize}
    \item \textbf{Environment module}: A Python class or module that loads the domain database and exposes tool functions with consistent interfaces.
    \item \textbf{Test suites}: Unit and workflow test files that serve as regression tests and documentation.
    \item \textbf{MCP server} (optional): A FastMCP server wrapping the environment with HTTP endpoints, session management, and transport options (stdio, HTTP, SSE), enabling direct integration with agent frameworks.
\end{itemize}
All artifacts are validated end to end before being passed to the DataAgent: the environment is instantiated against the database, tests are executed, and the tool interfaces are verified to match the specifications that the DataAgent will use for trajectory generation.

\subsubsection{Example Workflows}
\label{sec:coding-agent-workflows}

The EnhancedOrchestrator exposes the generate-test-debug loop described above through four high-level workflow modes, each tailored to a common development scenario:

\paragraph{Schema-Implement.}
Given a set of refined function schemas (signatures, docstrings, and behavioral specifications), the orchestrator drives the CodingAgent through a full implementation pass: the Planner decomposes the schemas into an ordered implementation plan respecting inter-function dependencies; the inner CodingAgent generates each function; the TestingAgent produces unit and workflow tests; and the iterative judge-based debug loop (Stage~1 then Stage~2) runs until all tests pass. The ReviewAgent performs a final quality gate before emitting the complete environment module. This is the primary mode for producing a new environment from scratch.

\paragraph{Code-Repair.}
When an existing implementation exhibits test failures or specification violations, the \emph{code-repair} workflow focuses the loop on diagnosis and repair. The orchestrator loads the current code and re-runs the test suites; for each failure, the JudgeAgent performs fault attribution to distinguish genuine code bugs from incorrect tests. The CodingAgent then applies targeted fixes---patching only the affected functions---while the debug loop ensures that repairs do not introduce regressions. 

\paragraph{Code-Patch.}
The \emph{code-patch} workflow handles incremental schema evolution: when function signatures, return types, or behavioral contracts are refined (e.g., a new optional parameter is added, or an error-handling convention is updated), the orchestrator diffs the old and new schemas, identifies the affected functions, and dispatches the CodingAgent in \emph{refine} mode to update only those implementations. The TestingAgent then regenerates or augments the affected tests to align with the updated schema, and the debug loop re-validates the patched code against both the original and revised specifications. This avoids costly full re-implementation when schemas undergo incremental refinement.

\paragraph{Env-to-MCP.}
The \emph{env-to-mcp} workflow converts an existing environment module into a standardized Model Context Protocol (MCP) server. The FileGenerationAgent wraps each tool function as a FastMCP endpoint with appropriate request/response schemas, session isolation, and transport configuration (stdio, HTTP, or SSE). The TestingAgent generates integration tests that exercise the MCP endpoints end to end, verifying that the server faithfully reproduces the behavior of the underlying environment. This mode enables seamless onboarding of environments into MCP-compatible agent frameworks without manual server authoring.

\paragraph{Format-Transform.}
The \emph{format-transform} workflow converts existing trajectory datasets into alternative tool-calling formats required by different model providers. Given a representative sample of the source dataset and a target format specification, the orchestrator dispatches the FileGenerationAgent and CodingAgent (inner) to synthesize a deterministic Python conversion script that maps dialogue turns, tool calls, and tool outputs into the target schema. The TestingAgent generates validation cases to ensure that the transformed outputs conform to the required structural conventions (e.g., field layout, argument nesting, and tool-call encoding). Once validated, the script is executed in batch across the dataset to produce the converted representation (e.g., OpenAI Harmony, Anthropic Messages API, Qwen, or DeepSeek), avoiding costly LLM-based rewriting while ensuring deterministic and structurally consistent conversions.
\subsection{DataAgent: Multi-Agent Trajectory Generation}
\label{sec:dataagent}

\begin{figure}[t]
    \centering
    \includegraphics[width=1.0\linewidth]{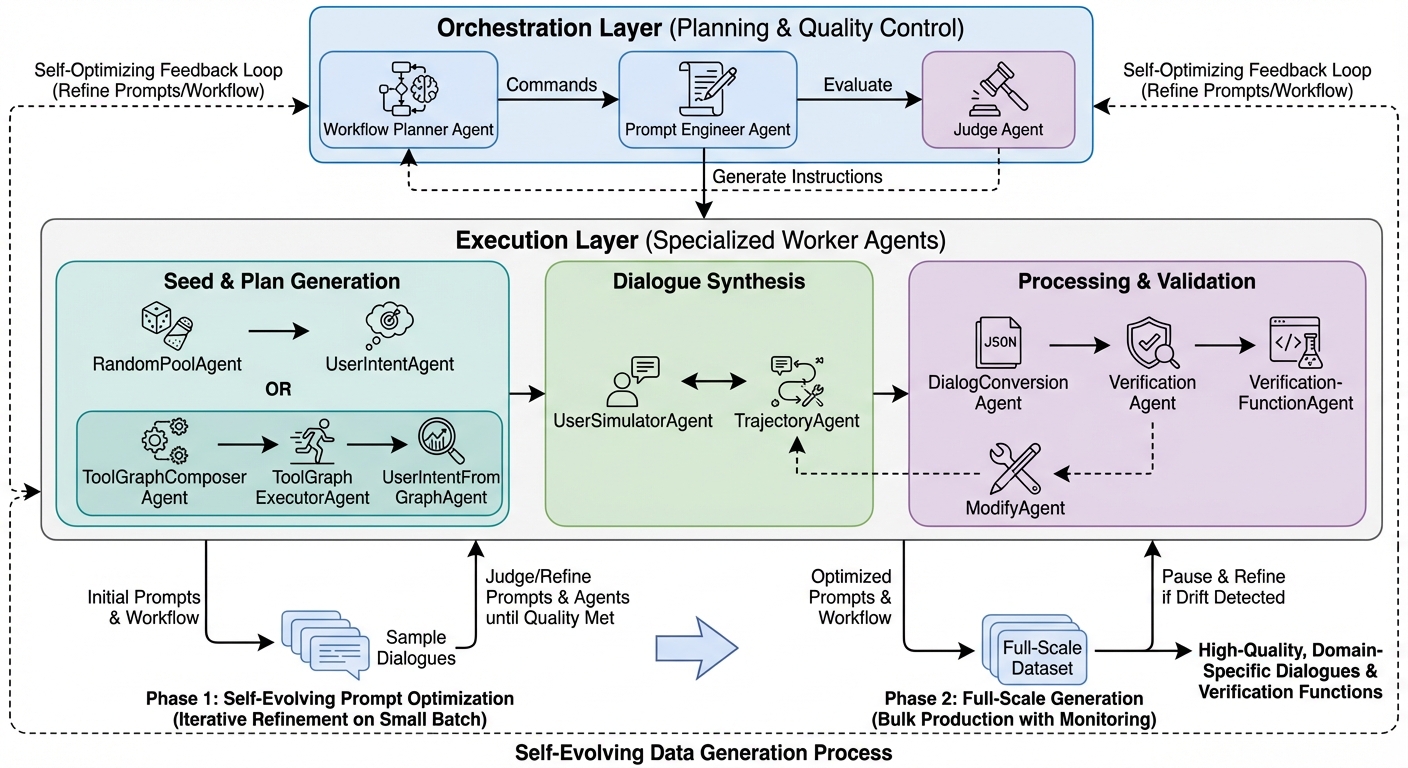}
    \caption{Internal architecture of the DataAgent. The Orchestration Layer (top) coordinates planning, prompt engineering, and quality evaluation through a self-optimizing feedback loop. The Execution Layer (bottom) comprises specialized worker agents organized into three stages: seed and plan generation, dialogue synthesis, and processing and validation. A two-phase self-evolving process first optimizes prompts on a small batch (Phase~1), then scales to full production with continuous monitoring (Phase~2).}
    \label{fig:dataagent}
\end{figure}

The DataAgent is the trajectory-generation engine of EigenData. Operating on the environments and databases produced by the DatabaseAgent and CodingAgent (or on existing artifacts for auditing and repair tasks), it orchestrates a hierarchy of specialized worker agents to synthesize, evaluate, and iteratively refine high-quality multi-turn function-calling data for both SFT and RL. The DataAgent comprises two internal layers: an \emph{Orchestration Layer} for planning, scheduling, and quality control, and an \emph{Execution Layer} of worker agents that perform domain-specific generation and analysis tasks.

\subsubsection{Orchestration Layer}
At the core of the DataAgent is an \textit{Orchestration Layer}, which coordinates the interactions among a set of high-level agents, guiding the generation process through iterative refine-and-evaluate cycles.

Drawing inspiration from recent hierarchical agent frameworks~\citep{liu2025hierarchicalmultiagentworkflowszeroshot,wong2025coachcollaborativeagentscontextual}, the Orchestrator first invokes a \textbf{Workflow Planner Agent} to devise a structured workflow as well as the models to be used in each part for data generation. The planner analyzes the user's request and domain specifications to plan an optimal generation workflow. It decides which execution agents to invoke, in what order and with which model, effectively decomposing the data generation task into a sequence of sub-tasks. The planner accounts for domain complexity and resource dependencies (ensuring, for example, that if a database query is needed, the corresponding agent is included). This results in a dynamic, context-aware workflow tailored to the target domain.

Next, a \textbf{Prompt Engineer Agent} is tasked with producing effective prompts for each worker agent in context. It auto-generates detailed instructions for the downstream agents, incorporating domain knowledge (e.g. domain guidelines, function specifications, or provided ontologies) and human requests. Crucially, the Prompt Engineer can revise its prompts over multiple rounds, informed by feedback from a \textbf{Judge Agent}, ensuring that instructions yield outputs meeting the desired accuracy, style, and compliance criteria.

The \textbf{Judge Agent} serves as an automated evaluator that critiques intermediate outputs along various dimensions -- including factual accuracy, relevance to the domain, coherence of conversation, adherence to formatting requirements, and overall realism. By leveraging large language models as judges~\citep{zheng2023judgingllmasajudge}, we obtain multi-dimensional quality assessments that have been shown to correlate well with human evaluations of dialog quality~\citep{yu2025aisjudgeaisrise}. The Judge's feedback is fed back into the loop, allowing the Orchestrator and Prompt Engineer to adjust subsequent prompts or agent behaviors. This forms a self-optimizing feedback loop (potentially augmented with human-in-the-loop oversight when available) in which the system ``self-verifies'' and improves its outputs each iteration. The orchestration layer therefore acts as a conductor, ensuring that all agents work in concert and that a high quality threshold is met before the process terminates.

\subsubsection{Execution Layer}
At the execution layer, EigenData orchestrates a sequence of specialized lower-level worker agents. Each agent is responsible for a distinct stage of the generation process, transforming the data step by step from abstract scenario seeds to finalized dialog records. In this work, we define a default collection of worker agents for the workflow planner to choose from, while preserving full modularity and extensibility. The DataAgent framework allows new agents to be added as needed, and the workflow planner dynamically configures the multi-agent architecture based on user requirements:
\begin{itemize}
    \item \textbf{RandomPoolAgent}: Generates a diverse pool of initial user scenarios and profiles. This agent creates varied starting conditions (user personas, goals, environmental contexts, constraint sets) to ensure broad coverage of possible dialogues. For example, in a customer service domain, RandomPoolAgent might output a range of customer types and inquiry topics as seeds for conversations.

    \item \textbf{UserIntentAgent}: Using a scenario from the pool (and any available domain data like databases or APIs), this agent synthesizes a concrete user intent for the dialogue. It may produce a detailed goal or task the user wants to accomplish, incorporating domain specifics. For instance, given a banking domain, the UserIntentAgent might formulate a user goal such as “dispute a credit card charge via the banking app,” including necessary details that will drive the ensuing conversation.

    \item \textbf{ToolGraphComposerAgent}: Given the available tools (APIs, functions, or services), this agent constructs a high-level tool-usage plan in the form of a compositional graph. The graph specifies the sequence, dependencies, and data flows among tools, ensuring that the overall workflow is both executable and aligned with domain constraints. It also performs sanity checks (e.g., missing parameters, incompatible outputs/inputs) and proposes alternative plans when conflicts arise.

    \item \textbf{ToolGraphExecutorAgent}: This agent executes the composed tool graph step-by-step, invoking tools, collecting intermediate results, and propagating outputs downstream. It monitors execution failures, retries with revised parameters when possible, and records structured traces of the tool-calling process. These traces serve both as supervision signals and as grounding evidence for the dialogue generation process.

    \item \textbf{UserIntentFromGraphAgent}: Using the executed tool graph and its traces, this agent reconstructs a coherent, user-centered narrative that explains why each tool call occurs and what the user is trying to achieve. It transforms system-level workflows into realistic, goal-driven user intents and sub-goals, ensuring that subsequent dialogues are grounded in actual executable behaviors rather than abstract high-level descriptions.

    \item \textbf{UserSimulatorAgent}: Simulates the user’s behavior and dialogue turns based on the user intent and persona. The UserSimulator generates realistic user utterances step-by-step, responding to the assistant and following the user’s goal. It maintains context and persona consistency (e.g. a polite customer vs. a frustrated one) and can handle increasing complexity over multi-turn interactions, ensuring the conversation flow remains natural from the user's side.

    \item \textbf{TrajectoryAgent}: Orchestrates the assistant’s side of the conversation, interacting with the UserSimulatorAgent to produce a full multi-turn multi-step function calling dialogue. The TrajectoryAgent uses the refined prompts and has access to domain tools or functions. If the domain provides an API or function (e.g., database lookup) that the assistant should call, the TrajectoryAgent will incorporate those function call actions into the dialogue at appropriate turns. This agent ensures the assistant’s responses are goal-directed and that any tool usage is executed correctly within the conversation. The output of this stage is a complete raw trajectory of the dialogue, including all user and assistant turns and any intermediate tool interactions.

    \item \textbf{DialogConversionAgent}: Converts the raw trajectory into a standardized dialogue format for the final dataset. This involves stripping out low-level execution details and formatting the conversation into the desired structure (for example, JSON or ChatML with clearly marked user/assistant messages, function call outputs, etc.). The DialogConversionAgent ensures that the final data is in an easily consumable format for model training or evaluation, including any necessary annotations (such as tags for function calls or system messages) as required by the target schema.

    \item \textbf{VerificationAgent}: Performs a thorough validation on the dialogue. The VerificationAgent checks dialogue quality and correctness one last time: it verifies that the conversation is coherent and logically consistent, that any domain-specific requirements are satisfied, and that all tool calls or API usages in the conversation have valid results. It also confirms that the formatting is correct and that no content violates the guidelines or domain rules. This automated verification (with dedicated checkers for tool outputs and content consistency) provides robust quality assurance, often catching issues that might be missed by human annotators. 

    \item \textbf{VerificationFunctionAgent}: Automatically constructs a programmatic verification function based on the dialogue trajectory and domain constraints. This function can be used for downstream evaluation or reinforcement learning as a reward model. It encodes correctness criteria, success conditions, tool-call validity, and failure signatures. For example, it may generate a scoring function that checks whether all required domain steps were completed, whether tool arguments align with constraints, and whether the user goal was achieved. This creates a self-contained evaluation artifact tied specifically to the generated dialogue instance.

    \item \textbf{ModifyAgent}: Applies targeted modifications to the dialogue based on external requests or issues detected by the VerificationAgent. It can fix formatting errors, rewrite inconsistent turns, adjust persona styles, regenerate problematic sections, or update tool calls that fail validation. This agent enables iterative refinement until the dialogue satisfies all quality, schema, and domain criteria, effectively closing the loop in a self-improving data generation pipeline.

    \item \textbf{DialogAnalyzerAgent}: Given an existing dialogue and a set of function schema changes, this agent identifies which conversation turns require regeneration and which can be preserved. It algorithmically segments the conversation into \textit{safe zones} (turns prior to the first affected call, replayed verbatim), \textit{affected turns} (turns that invoke a renamed or schema-mismatched function), and \textit{candidate zones} (turns following an affected call that may depend on the now-invalidated tool output); dialogues that contain no affected function calls are marked for passthrough, bypassing all downstream agents. For affected dialogues, the agent additionally invokes the LLM to extract the user's intent and produce continuation guidance that enables the TrajectoryAgent to resume the conversation coherently from the truncation point. 
\end{itemize}

\subsubsection{Self-Evolving Data Generation Process}
\label{sec:eigendata-two-phase}

To balance quality and efficiency, the DataAgent operates in a two-phase generation regime.
\paragraph{Phase~1: Self-Evolving Prompt Optimization.}
The orchestrator generates a small batch of sample dialogues (e.g. 5–20) using initial prompts, then engages the Judge and other high-level agents in iterative refinement on this batch. Through a few iterations, the prompts and agent behaviors are tuned until the sample dialogues satisfy the quality criteria. This corresponds to an initial coarse-to-fine optimization of the prompt and workflow, analogous to recent self-referential prompt evolution~\citep{fernando2023promptbreeder} and declarative pipeline compilation~\citep{khattab2023dspy} methods that first refine prompts then produce final answers.

\paragraph{Phase~2: Full-Scale Generation.} Once the prompt strategy is optimized, the orchestrator uses the refined prompts to generate the full dataset at scale. The Execution Layer are run to produce the complete set of dialogues, while monitoring mechanisms (including the Judge and VerificationAgent) continue to evaluate quality on the fly. If any drift or new error pattern is observed during bulk generation, the system can pause and invoke the refinement loop again to address it. After generation, a final validation pass is applied by the VerificationAgent to ensure all dialogues meet the release standards.

This two-phase approach focuses human or computational critique on a small pilot batch to avoid large-scale errors, then confidently expands to the entire data request. By integrating hierarchical control, specialized agents, and iterative self-evolving, the DataAgent is able to automatically synthesize high-quality, realistic conversational datasets. The general-purpose design, with pluggable domain resources and no task-specific training required, allows it to be readily adapted to a wide range of domains and dialogue styles from open-domain chat to task-oriented dialogs, while consistently maintaining rigorous quality control throughout the data generation process.

\subsubsection{Example Workflows}
\label{sec:example-workflows}

\begin{figure}[t]
\centering
\includegraphics[width=1.0\linewidth]{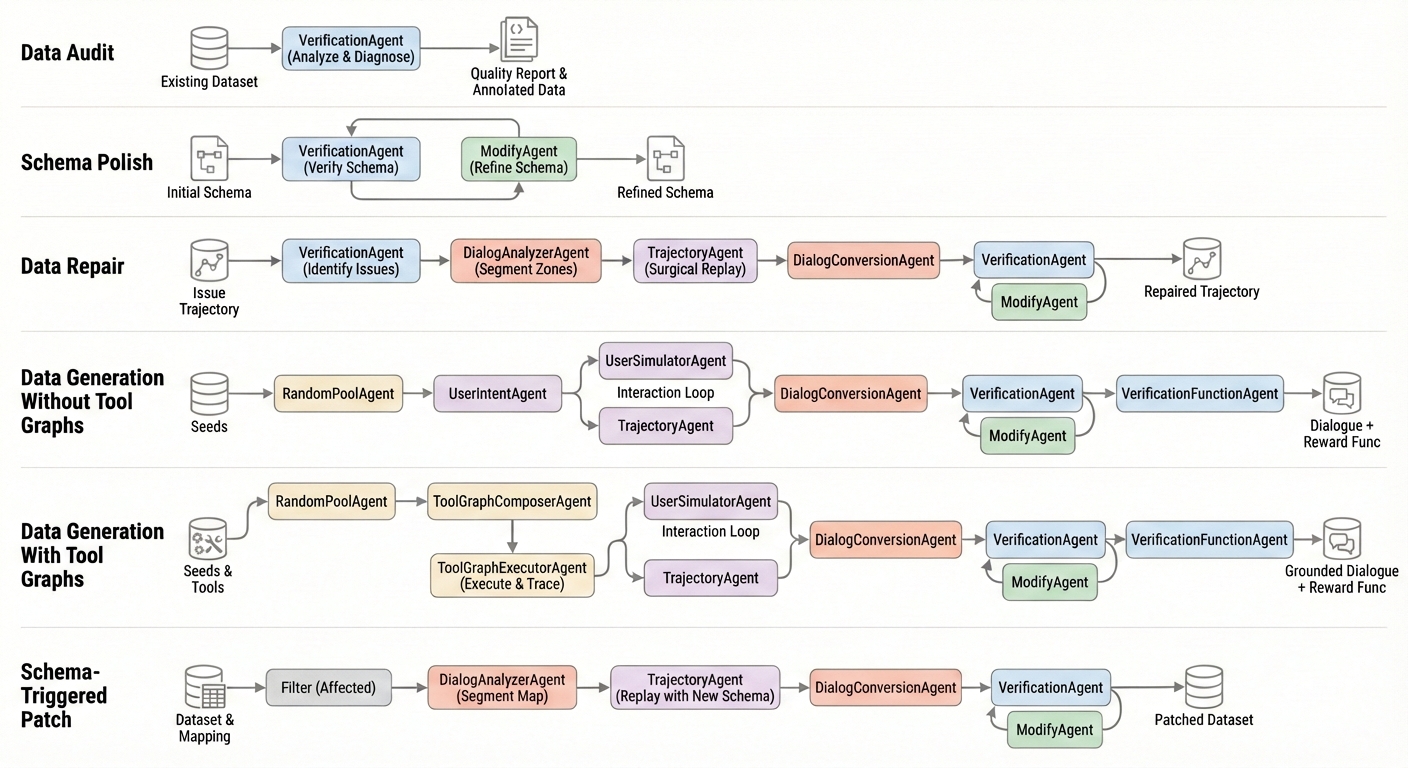}
\caption{Representative workflows instantiated by the DataAgent's Workflow Planner. Each row shows a distinct pipeline: \textbf{Data Audit} systematically diagnoses quality issues and failure modes; \textbf{Schema Polish} iteratively refines API specifications via verification--modification loops; \textbf{Data Repair} performs targeted trajectory repair by segmenting conversations into safe, affected, and candidate zones; \textbf{Data Generation Without Tool Graphs} produces multi-turn conversations top-down from sampled user intents; \textbf{Data Generation With Tool Graphs} generates grounded conversations bottom-up from executed tool-call plans; \textbf{Schema-Triggered Patch} incrementally reconstructs only the trajectory portions affected by schema evolution.}
\label{fig:workflows}
\end{figure}

We now illustrate several representative workflows that the DataAgent's high-level planner may instantiate for different data-generation objectives (see Figure~\ref{fig:workflows}).

\paragraph{Data Audit.}
This workflow systematically audits an existing dataset to expose quality issues, coverage gaps, and model failure modes. The \textit{VerificationAgent} inspects each trajectory against the function schema and domain business rules, going beyond a binary pass/fail decision to perform structured error analysis: it identifies (i)~turns that violate domain constraints, (ii)~incorrect or unsafe tool usage, (iii)~unmet user goals, and (iv)~contradictions or hallucinations. In parallel, it analyzes the distribution and coverage of functions used across the dataset compared to the full list of available functions. The output is a structured report that annotates individual trajectories with richly labeled diagnostics and summarizes dataset-level issues, anomalies, coverage gaps, and quality metrics---directly useful for dataset debugging, reliability benchmarking, and robustness training.

\paragraph{Schema Polish.}
In domains with underspecified or evolving APIs, EigenData supports iterative refinement of tool schemas. When the \textit{VerificationAgent} detects recurring errors in function calls (e.g., missing arguments, ambiguous parameter semantics, or incorrect output assumptions), it attributes the failure to specific schema deficiencies. These structured diagnostics are then passed to the \textit{ModifyAgent}, which rewrites or augments the schema—adding constraints, clarifying parameter descriptions, or documenting edge cases. The updated schema is reinserted into the pipeline and revalidated, gradually converging toward executable, unambiguous specifications.

\paragraph{Data Repair.}
When trajectories are known to contain quality issues but the specific problematic turns are not identified in advance, EigenData applies targeted repair rather than regenerating entire conversations. The \textit{VerificationAgent} first performs a structured analysis of the dialogue, producing a list of problematic turn indices that flag where domain constraints are violated, tool calls are incorrect, or reasoning is inconsistent (e.g., an assistant turn that invokes a function with invalid arguments or a response that contradicts prior context). A \textit{DialogAnalyzerAgent} then uses these indices to segment the conversation into \textit{safe zones} (unaffected turns, preserved verbatim), \textit{affected turns} (turns at the flagged indices that must be regenerated), and \textit{candidate zones} (turns following an affected turn whose validity may depend on the repaired output). The \textit{TrajectoryAgent} performs a surgical replay, threading safe-zone turns through unchanged while regenerating affected and candidate turns against the live tool environment. The repaired dialogue then passes through the standard \textit{DialogConversionAgent}--\textit{VerificationAgent}--\textit{ModifyAgent} pipeline for format compliance and a final consistency check, yielding polished trajectories with minimal deviation from otherwise valid conversations.

\paragraph{Data Generation Without Tool Graphs.}
For open-ended or conversational domains, EigenData operates in a primarily top-down mode. The \textit{RandomPoolAgent} first samples diverse scenario factors (personas, goals, constraints, contexts), and the \textit{UserIntentAgent} instantiates a concrete task from one scenario. Dialogue is then generated through interaction between the \textit{UserSimulatorAgent} and the \textit{TrajectoryAgent}, producing a realistic multi-turn exchange. The \textit{DialogConversionAgent} formats the conversation, while the \textit{VerificationAgent} ensures coherence, safety, and formatting compliance. Minor issues are corrected by the \textit{ModifyAgent}. Finally, the \textit{VerificationFunctionAgent} synthesizes an executable reward/verification function, enabling reinforcement learning and evaluation on the generated instance.

\paragraph{Data Generation With Tool Graphs.}
For goal-oriented domains with executable tools, EigenData generates grounded, bottom-up conversations. The \textit{RandomPoolAgent} samples user profiles and contexts. The \textit{ToolGraphComposerAgent} constructs a compositional plan over available APIs, and the \textit{ToolGraphExecutorAgent} executes the plan, completing parameters, handling failures, and recording traces. The \textit{UserIntentFromGraphAgent} converts these traces into a coherent, goal-driven narrative aligned with the executed workflow. The \textit{UserSimulatorAgent} and \textit{TrajectoryAgent} then generate the full dialog, where tool interactions emerge naturally during conversation. The \textit{DialogConversionAgent} standardizes the trajectory, while the \textit{VerificationAgent} validates logical consistency, tool correctness, and formatting. If necessary, the \textit{ModifyAgent} performs selective repairs. Finally, the \textit{VerificationFunctionAgent} constructs a programmatic checker tied to the underlying tool graph, enabling precise, reproducible evaluation.

\textbf{Schema-Triggered Patch.} When a deployed function schema evolves---for instance, a function is renamed, a new required parameter is introduced, or an argument type is revised---existing trajectories that invoke the affected functions become misaligned with the updated specification. Rather than discarding and regenerating the entire dataset, the \textit{schema-triggered patch} workflow applies targeted reconstruction at the trajectory level, confining regeneration cost strictly to the affected portions of the corpus. The orchestrator accepts a user-supplied \textit{function mapping} (e.g., renaming \texttt{cancel\_reservation} to \texttt{cancel\_booking}) and first filters the full dataset to identify only the dialogues that contain calls to the modified functions, forwarding all others unchanged. For each affected dialogue, a \textit{DialogAnalyzerAgent} decomposes the conversation into a structured \textit{segment map} that classifies turns as a \textit{safe zone} (turns prior to the first affected call, replayed verbatim), an \textit{affected turn} (turns that directly invoke the changed function or call it with arguments that violate the updated schema), or a \textit{candidate zone} (turns following an affected call whose content may depend on the now-invalidated tool output). The \textit{TrajectoryAgent} then performs a surgical replay, threading safe-zone turns through unchanged while regenerating affected and candidate turns against the live tool environment using the updated schema. The repaired trajectories re-enter the standard \textit{DialogConversionAgent}--\textit{VerificationAgent}--\textit{ModifyAgent} pipeline for format compliance and consistency checks, making schema evolution an incremental, cost-proportional operation rather than a full-pipeline restart.

\subsection{End-to-End Generation Workflow}
\label{sec:e2e-workflow}

We now describe the canonical end-to-end workflow for generating a complete training environment and dataset from a domain specification---the pipeline that EigenCore orchestrates when a user requests from-scratch generation.

\paragraph{Step~1: Database Construction (DatabaseAgent).}
Given the domain specification, the DatabaseAgent designs the database schema, populates it with realistic synthetic records, and validates internal consistency. The output is a self-contained database artifact (e.g., a JSON file or SQLite database) together with its formal schema.

\paragraph{Step~2: Environment Generation (CodingAgent).}
The CodingAgent receives the database and domain specification and generates the executable environment: a Python module (or MCP server) that exposes tool functions backed by the database. It iteratively tests and debugs the implementation until all unit and workflow tests pass, producing a verified environment along with its test suite and function schemas.

\paragraph{Step~3: Trajectory Synthesis (DataAgent).}
The DataAgent loads the verified environment and database, and uses its hierarchical multi-agent pipeline to generate diverse, high-quality multi-turn trajectories. Through its self-evolving two-phase process (Section~\ref{sec:eigendata-two-phase}), it first optimizes prompts on a small pilot batch, then scales to full production with continuous quality monitoring.

\paragraph{Cross-Component Feedback.}
Throughout the pipeline, EigenCore monitors for cross-component inconsistencies. If the DataAgent's VerificationAgent detects that a tool function behaves incorrectly (e.g., returns unexpected results for valid inputs), EigenCore can route the issue back to the CodingAgent for targeted repair. Similarly, if the CodingAgent discovers that the database lacks records needed for meaningful tool-use scenarios, EigenCore can request the DatabaseAgent to augment the data. This feedback mechanism ensures that the three sub-systems converge toward a consistent, high-quality output without requiring manual intervention.

\section{Benchmark Auditing, Repair, and Extension: A Case Study on BFCL}
BFCL (Berkeley Function-Calling Leaderboard)~\citep{patilberkeley} is one of the most widely adopted benchmarks for evaluating large language models' function-calling capabilities. It assesses a model's ability to correctly select functions, generate valid arguments, and conduct multi-step tool calling across diverse domains. Given its prominence as a standard evaluation suite for tool-use reliability and compositional reasoning in LLMs~\citep{patilberkeley}, BFCL serves as a representative case study for examining the quality of existing benchmarks and demonstrating how \textsc{EigenData} can be applied to audit, repair, and extend them.

In this case study, \textsc{EigenCore} (Section~\ref{sec:eigencore}) orchestrates the full audit-and-repair pipeline, dispatching sub-tasks to the appropriate sub-systems: the \textsc{DataAgent} (Section~\ref{sec:dataagent}) drives error analysis and trajectory repair through its \textit{VerificationAgent} and \textit{ModifyAgent}; the \textsc{CodingAgent} (Section~\ref{sec:coding-agent}) diagnoses and fixes implementation bugs in the benchmark's executable functions; and the \textsc{DatabaseAgent} (Section~\ref{sec:database-agent}) validates and corrects any underlying data inconsistencies. Cross-component feedback (Section~\ref{sec:e2e-workflow}) ensures that fixes in one layer propagate to all dependent artifacts.

We focus specifically on the \textbf{multi-turn} subset of BFCL, which evaluates multi-step tool-use trajectories requiring sequential reasoning and state tracking---the most challenging and practically relevant setting for real-world agent deployment.

\subsection{Error within BFCL}
To audit the quality of BFCL's multi-turn evaluation suite, \textsc{EigenCore} dispatches a coordinated analysis across all three sub-systems. The \textsc{DataAgent}'s \textit{VerificationAgent} examines ground-truth trajectories and evaluation logic via the \emph{Data Audit} workflow (Section~\ref{sec:dataagent}); the \textsc{CodingAgent}'s \textit{TestingAgent} and \textit{JudgeAgent} exercise the benchmark's executable functions against synthesized edge cases (Section~\ref{sec:coding-agent}); and the \textsc{DatabaseAgent}'s \textit{ConsistencyVerificationAgent} cross-checks backing data for constraint violations (Section~\ref{sec:database-agent}). This multi-agent audit reveals three categories of errors that compromise the reliability of BFCL as a benchmark.

\paragraph{Function Schema Errors (affecting 39 entries, 19.5\%).}
We identify 39 entries where the function schemas provided to models contain incorrect or incomplete specifications. These include: (1)~\emph{parameter type mismatches}, where a parameter is declared as one type but the reference implementation or ground truth expects another---for instance, \texttt{close\_ticket} declares \texttt{ticket\_id} as \texttt{integer}, yet the initial configuration stores ticket IDs as strings and the ground truth passes the string \texttt{"ticket\_001"} (Figure~\ref{fig:bfcl-schema-error}); (2)~\emph{missing required parameters}, where the ground-truth trajectory assumes the model should infer values without any discovery mechanism; and (3)~\emph{ambiguous or contradictory parameter descriptions} that admit multiple valid interpretations---such as \texttt{setCruiseControl} using \texttt{"m/h"} for input but \texttt{"km/h"} for output (Figure~\ref{fig:bfcl-schema-unit}), or \texttt{get\_order\_history} documenting a response key \texttt{"order\_history"} while the implementation returns \texttt{"history"} (Figure~\ref{fig:bfcl-schema-response}). Appendix~\ref{sec:bfcl-error-examples} provides additional examples.

\paragraph{Function Implementation Errors (affecting 91 entries, 45.5\%).}
The executable functions backing BFCL's evaluation contain implementation bugs that cause correct model outputs to be scored as failures. We find 91 affected entries spanning three recurring patterns: (1)~\emph{logic errors}, such as the \texttt{setHeadlights} function where the \texttt{"auto"} mode---a schema-valid enum value---falls into the \texttt{else} branch and silently turns headlights \emph{off} instead of providing automatic behavior (Figure~\ref{fig:bfcl-impl-error}), or \texttt{tail(lines=0)} returning the \emph{entire} file instead of nothing due to Python's \texttt{content[-0:]} semantics (Figure~\ref{fig:bfcl-impl-tail}); (2)~\emph{input handling defects}, including \texttt{get\_zipcode\_based\_on\_city} which silently returns an invalid zipcode \texttt{"00000"} for unrecognized or differently-cased city names instead of raising an error (Figure~\ref{fig:bfcl-impl-zipcode}); and (3)~\emph{state management bugs}, such as \texttt{send\_message} storing messages without a sender field, making it impossible to distinguish senders when deleting or viewing messages (Figure~\ref{fig:bfcl-impl-sendmsg}). Because the evaluation pipeline executes these functions to determine correctness, these bugs silently penalize models that produce valid outputs, inflating error rates and distorting cross-model comparisons.

\begin{figure}[htbp]
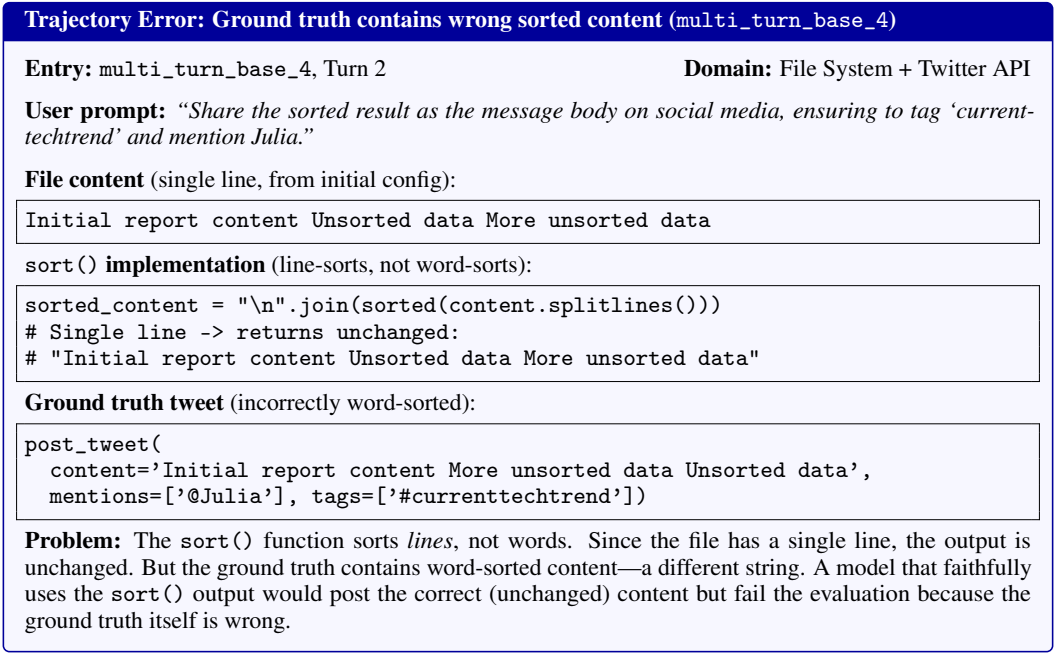

\centering
\begin{tcolorbox}[
  colback=blue!3,
  colframe=blue!60!black,
  fonttitle=\bfseries\small,
  title={Trajectory Error: Ground truth contains wrong sorted content (\texttt{multi\_turn\_base\_4})},
  boxrule=0.6pt,
  arc=2pt,
  left=5pt, right=5pt, top=4pt, bottom=4pt
]
\small
\textbf{Entry:} \texttt{multi\_turn\_base\_4}, Turn 2 \hfill \textbf{Domain:} File System + Twitter API

\medskip
\textbf{User prompt:} \textit{``Share the sorted result as the message body on social media, ensuring to tag `currenttechtrend' and mention Julia.''}

\medskip
\textbf{File content} (single line, from initial config):
\begin{lstlisting}[style=jsonstyle, aboveskip=4pt, belowskip=4pt]
Initial report content Unsorted data More unsorted data
\end{lstlisting}

\textbf{\texttt{sort()} implementation} (line-sorts, not word-sorts):
\begin{lstlisting}[style=jsonstyle, aboveskip=4pt, belowskip=4pt]
sorted_content = "\n".join(sorted(content.splitlines()))
# Single line -> returns unchanged:
# "Initial report content Unsorted data More unsorted data"
\end{lstlisting}

\textbf{Ground truth tweet} (incorrectly word-sorted):
\begin{lstlisting}[style=jsonstyle, aboveskip=4pt, belowskip=4pt]
post_tweet(
  content='Initial report content More unsorted data Unsorted data',
  mentions=['@Julia'], tags=['#currenttechtrend'])
\end{lstlisting}

\textbf{Problem:} The \texttt{sort()} function sorts \emph{lines}, not words. Since the file has a single line, the output is unchanged. But the ground truth contains word-sorted content---a different string. A model that faithfully uses the \texttt{sort()} output would post the correct (unchanged) content but fail the evaluation because the ground truth itself is wrong.
\end{tcolorbox}
\caption{Example of a trajectory error in BFCL. The ground truth contains word-sorted content, but the \texttt{sort()} function sorts lines---producing unchanged output for a single-line file. The reference answer is provably incorrect.}
\label{fig:bfcl-traj-error}
\end{figure}

\paragraph{Trajectory and Evaluation Errors (affecting 82 entries, 41.9\%).}
The ground-truth trajectories used as references exhibit 82 errors across three sub-categories: (1)~\emph{overly rigid evaluation}, which penalizes functionally or semantically equivalent but syntactically different function call sequences---for example, rename-then-move versus the ground truth's move-then-rename, despite producing identical filesystem states (Figure~\ref{fig:bfcl-traj-ordering}), or requiring integer-formatted tire pressures (\texttt{32}) when the API returns floats (\texttt{32.0}) (Figure~\ref{fig:bfcl-traj-float}); (2)~\emph{incorrect ground-truth labels}, where the reference trajectory itself contains erroneous outputs, such as a ground truth that word-sorts a single-line file instead of line-sorting it (Figure~\ref{fig:bfcl-traj-error}), or \texttt{du()} being called with \texttt{human\_readable=True} but the ground truth writing the non-human-readable format (Figure~\ref{fig:bfcl-traj-du}); and (3)~\emph{missing alternative solution paths}, where a model that achieves the correct final state through a different but valid sequence of calls is marked as incorrect.

In total, these errors affect \textbf{143 out of 200} entries (\textbf{71.5\%}) in BFCL's multi-turn evaluation suite, undermining the reliability of model rankings derived from the original benchmark. Table~\ref{tab:error-summary} summarizes the error distribution.

\begin{table}[htbp]
\centering
\caption{Summary of errors identified in BFCL multi-turn base evaluation suite (200 entries).}
\label{tab:error-summary}
\begin{tabular}{lcc}
\toprule
\textbf{Error Category} & \textbf{Entries Affected} & \textbf{\% of Cases} \\
\midrule
Function Schema Errors        & 39  & 19.5\%  \\
Function Implementation Errors          & 91  & 45.5\%  \\
Trajectory / Evaluation Errors & 82  & 41.0\%  \\
\midrule
\textbf{Total (deduplicated)}  & \textbf{143}  & \textbf{71.5\%}  \\
\bottomrule
\end{tabular}
\end{table}


\begin{figure}[htbp]
\centering
\begin{tcolorbox}[
  colback=red!3,
  colframe=red!60!black,
  fonttitle=\bfseries\small,
  title={Schema Error: Schema declares \texttt{integer}, ground truth passes \texttt{string} (\texttt{close\_ticket})},
  boxrule=0.6pt,
  arc=2pt,
  left=5pt, right=5pt, top=4pt, bottom=4pt
]
\small
\textbf{Entry:} \texttt{multi\_turn\_base\_173}, Turn 3 \hfill \textbf{Domain:} Ticket API

\medskip
\textbf{User prompt:} \textit{``I noticed there's a ticket linked with this booking that I no longer find necessary. Can you cancel it on my behalf?''}

\medskip
\textbf{Schema definition} (\texttt{ticket\_api.json}):
\begin{lstlisting}[style=jsonstyle, aboveskip=4pt, belowskip=4pt]
{ "name": "close_ticket",
  "parameters": {
    "ticket_id": {"type": "integer",
                  "description": "ID of the ticket to be closed."}
}}
\end{lstlisting}

\textbf{Ground truth call:}
\begin{lstlisting}[style=jsonstyle, aboveskip=4pt, belowskip=4pt]
close_ticket(ticket_id='ticket_001')   # passes a string!
\end{lstlisting}

\textbf{Initial config} (ticket IDs stored as strings):
\begin{lstlisting}[style=jsonstyle, aboveskip=4pt, belowskip=4pt]
{"ticket_queue": [{"id": "ticket_001", "status": "Open", ...}]}
\end{lstlisting}

\textbf{Problem:} The schema specifies \texttt{ticket\_id} as \texttt{integer}, but the initial config stores ticket IDs as the string \texttt{"ticket\_001"} and the ground truth passes a string. The implementation's lookup (\texttt{ticket["id"] == ticket\_id}) compares by equality, so passing an integer---as the schema dictates---would never match the string key. A model that follows the schema strictly produces a type-correct call that fails at runtime.
\end{tcolorbox}
\caption{Example of a function schema error in BFCL. The schema declares \texttt{ticket\_id} as \texttt{integer}, but the ground truth and backing data use the string \texttt{"ticket\_001"}---a model following the schema cannot succeed.}
\label{fig:bfcl-schema-error}
\end{figure}

\begin{figure}[htbp]
\centering
\begin{tcolorbox}[
  colback=orange!3,
  colframe=orange!70!black,
  fonttitle=\bfseries\small,
  title={Implementation Error: \texttt{setHeadlights("auto")} silently turns headlights OFF},
  boxrule=0.6pt,
  arc=2pt,
  left=5pt, right=5pt, top=4pt, bottom=4pt
]
\small
\textbf{Entry:} \texttt{multi\_turn\_base\_50} \hfill \textbf{File:} \texttt{vehicle\_control.py:314--329}

\medskip
\textbf{User prompt:} \textit{``It's getting a bit darker out here, so please flip on the headlights for visibility.''}

\medskip
\textbf{Implementation:}
\begin{lstlisting}[style=jsonstyle, aboveskip=4pt, belowskip=4pt]
def setHeadlights(self, mode: str):
    if mode not in ["on", "off", "auto"]:
        return {"error": "Invalid headlight mode."}
    if mode == "on":
        self.headLightStatus = "on"
        return {"headlightStatus": "on"}
    else:  # Both "off" AND "auto" fall here!
        self.headLightStatus = "off"
        return {"headlightStatus": "off"}
\end{lstlisting}

\textbf{Problem:} The schema declares \texttt{"auto"} as a valid enum value, but the \texttt{else} branch treats both \texttt{"off"} and \texttt{"auto"} identically---turning headlights \emph{off}. A model that reasonably interprets darkening conditions and calls \texttt{setHeadlights(mode="auto")} would get \texttt{headlightStatus="off"} instead of \texttt{"on"}, causing a state mismatch and test failure. The ground truth avoids this by calling \texttt{setHeadlights(mode="on")}, masking the bug.
\end{tcolorbox}
\caption{Example of a function implementation error in BFCL. The \texttt{"auto"} headlight mode---a schema-valid option---silently falls into the \texttt{"off"} branch due to a missing \texttt{elif}, penalizing models that make a reasonable choice.}
\label{fig:bfcl-impl-error}
\end{figure}

\subsection{Benchmark Fix with EigenData}
\label{sec:bfcl-fix}

\textsc{EigenCore} orchestrates the repair pipeline by routing each error category to the appropriate sub-system and enforcing cross-component consistency through its feedback loop (Section~\ref{sec:e2e-workflow}). Schema errors are handled by the \textsc{DataAgent}'s schema-polish workflow, implementation bugs by the \textsc{CodingAgent}'s iterative debug loop, and trajectory errors by the \textsc{DataAgent}'s modification workflow---with \textsc{EigenCore} ensuring that a schema fix propagates to both the corresponding implementation and all affected trajectories. We select 200 representative multi-turn cases (BFCL-v3 set) for detailed repair and re-evaluation.

\paragraph{Function Schema Repair.}
We apply \textsc{EigenData}'s \emph{Schema Polish} workflow (Section~\ref{sec:example-workflows}) to systematically correct deficient schemas. The \textit{VerificationAgent} first cross-references each function schema against its reference implementation and ground-truth trajectory usage, detecting recurring mismatches such as parameter type conflicts, missing constraints, and ambiguous descriptions. These structured diagnostics are passed to the \textit{ModifyAgent}, which rewrites the schema---correcting type declarations, adding missing parameter constraints, and clarifying descriptions to resolve ambiguity. The updated schema is reinserted into the pipeline and revalidated by the \textit{VerificationAgent}, iterating until the specification is consistent with both its implementation and its downstream usage in trajectories.

\paragraph{Function Implementation Repair.}
For implementation bugs, we leverage \textsc{EigenData}'s \emph{CodingAgent} (Section~\ref{sec:coding-agent}) to diagnose and correct the faulty executable functions. The repair proceeds in three stages. First, the \textit{TestingAgent} executes the existing test suite and collects structured failure traces---stack traces, assertion mismatches, and input--output pairs that trigger incorrect behavior. Second, the \textit{JudgeAgent} analyzes these traces to attribute each failure to either a genuine implementation bug or a test-side issue (e.g., an overly strict assertion or an incorrect expected value), preventing repairs from chasing false positives. Third, for confirmed implementation bugs, the \textit{FileGenerationAgent} rewrites the defective function body while preserving the corrected schema contract established in the schema-repair step. Each rewritten function is revalidated by the \textit{TestingAgent} against both the original test cases and additional edge cases synthesized by the pipeline---including boundary inputs, null/empty arguments, and adversarial orderings identified during the error-analysis phase. This iterative debug loop (up to $M$ iterations per cycle, $N$ cycles per function) continues until all identified implementation bugs are resolved, yielding regression coverage beyond what the original benchmark's fixed test cases offer.

\paragraph{Trajectory and Evaluation Repair.}
We employ \textsc{EigenData}'s \emph{Data Repair} workflow to repair ground-truth trajectories. The \textit{VerificationAgent} scans each reference trajectory and marks spans containing incorrect function calls, invalid argument values, or logically inconsistent reasoning. Rather than regenerating entire conversations, the \textit{ModifyAgent} rewrites only the problematic segments, preserving valid reasoning and context. The repaired trajectory is then rechecked through a verification--modification--verification loop until all issues are resolved. For evaluation, the \textit{VerificationFunctionAgent} synthesizes state-based checking functions that assess whether the final system state matches the expected outcome, rather than requiring exact trajectory matching. This allows the evaluation to accept functionally equivalent call sequences, and for cases where multiple valid solution paths exist, the verification function encodes all acceptable completion criteria.

\subsection{Evaluation Results}
\label{sec:bfcl-results}

We evaluate five frontier models on the repaired 200-case subset (BFCL-v3) using both automated metrics and human evaluation to assess the consistency and reliability of the corrected benchmark.

\paragraph{Automated Evaluation.}
Table~\ref{tab:auto-eval} reports results across five automated metrics of increasing strictness: \emph{Config Match} checks whether the model's function call configuration (function name and parameter schema) matches the reference; \emph{Key Function} verifies that critical function calls are present; \emph{LLM-as-Judge} uses a separate language model to assess trajectory correctness; \emph{Config$+$Func} requires both configuration and key function match; and \emph{All Three} demands simultaneous satisfaction of configuration match, key function match, and LLM-as-Judge approval.

\begin{table}[htbp]
\centering
\caption{Automated evaluation results on the repaired BFCL multi-turn subset (200 cases). Models are sorted by the strictest metric (All Three). Best results per column are \textbf{bolded}.}
\label{tab:auto-eval}
\renewcommand{\arraystretch}{1.15}
\setlength{\tabcolsep}{5pt}
\begin{tabular}{l
  >{\centering\arraybackslash}p{1.8cm}
  >{\centering\arraybackslash}p{1.8cm}
  >{\centering\arraybackslash}p{1.8cm}
  >{\centering\arraybackslash}p{1.8cm}
  >{\centering\arraybackslash}p{1.8cm}}
\toprule
\textbf{Model}
  & \textbf{Config Match}
  & \textbf{Key Func.}
  & \textbf{LLM Judge}
  & \textbf{Config $+$ Func}
  & \textbf{All Three} \\
\midrule
\rowcolor{bestrow}
Gemini-3-Pro        & \textbf{65.5} & \textbf{71.0} & \textbf{59.0} & \textbf{53.0} & \textbf{42.5} \\
Claude-4.5-Sonnet   & 57.5          & 57.5          & 52.0          & 43.0          & 38.0          \\
\rowcolor{lightgray}
GLM-4.6             & 50.5          & 58.5          & 49.0          & 39.5          & 32.5          \\
GPT-5.2             & 50.5          & 54.0          & 45.0          & 40.0          & 32.5          \\
\rowcolor{lightgray}
DeepSeek-V3.2       & 45.0          & 44.0          & 40.0          & 29.5          & 22.5          \\
\midrule
\textit{Average}    & \textit{53.8} & \textit{57.0} & \textit{49.0} & \textit{41.0} & \textit{33.6} \\
\bottomrule
\end{tabular}
\vspace{2pt}
\end{table}

\paragraph{Human Evaluation.}

To assess the reliability of automated metrics and further validate the corrected benchmark, we conduct a human evaluation study on five representative models, collecting approximately 40 evaluation cases per model.

To construct realistic evaluation scenarios, annotators first select several target environments and optionally configure complex settings (e.g., API availability, tool configurations, or multi-tool environments). These configurations initialize the evaluation context. An LLM-assisted interface then supports annotators in generating diverse and challenging multi-turn user queries: given the selected environment and configuration, the LLM proposes candidate task descriptions and follow-up interactions, which annotators refine or extend to produce realistic multi-step requests. This process enables the creation of complex agentic tasks involving tool usage, state tracking, and iterative reasoning. For each generated task, the same prompt sequence is executed across all evaluated models. Annotators are presented with the resulting trajectories and final outputs from each model and determine whether the model successfully fulfills the user's intent under the given environment. Each model output is labeled as \textbf{Pass} or \textbf{Fail}: a \textbf{Pass} indicates that the model correctly uses available tools to produce a functionally valid solution satisfying the user's request, regardless of differences in intermediate reasoning steps or tool invocation order; a \textbf{Fail} indicates that the model produces an incorrect result, an incomplete solution, or invalid tool interactions that prevent successful task completion.

Crucially, unlike automated evaluation, which typically relies on trajectory matching against predefined references, our human evaluation does not use reference trajectories at all. Annotators judge task success based solely on whether the model correctly uses available tools to accomplish the user's intent within the given environment. This reference-free design captures valid alternative reasoning paths and tool usage strategies that automated trajectory-based metrics systematically miss, providing a more faithful measure of practical tool-use capability in realistic multi-turn scenarios. Table~\ref{tab:human-eval} reports the results alongside each model's ranking on the original BFCL multi-turn leaderboard.\footnote{https://gorilla.cs.berkeley.edu/leaderboard.html}

\begin{table}[htbp]
\centering
\caption{Human evaluation results on the repaired BFCL environment. Annotators judge whether each model trajectory constitutes a functionally correct task completion without reference trajectories. Pass rates are compared against the strictest automated metric (All Three) on the repaired benchmark and against the original BFCL multi-turn leaderboard scores. Models are sorted by human pass rate.}
\label{tab:human-eval}
\renewcommand{\arraystretch}{1.15}
\setlength{\tabcolsep}{6pt}
\begin{tabular}{l c c c c c}
\toprule
\textbf{Model}
  & \textbf{Pass}
  & \textbf{Fail}
  & \textbf{Pass Rate (\%)}
  & \makecell{\textbf{Repaired BFCL}\\\textbf{All Three (\%)}}
  & \makecell{\textbf{Original}\\\textbf{BFCL MT (\%)}} \\
\midrule
\rowcolor{bestrow}
Gemini-3-Pro        & 33 & 8 & \textbf{80.5} & \textbf{42.5} & 63.12 \\
Claude-4.5-Sonnet   & 28 & 13  & 68.3          & 38.0          & 61.37 \\
\rowcolor{lightgray}
GPT-5.2             & 26 & 14  & 65.0          & 32.5          & 28.12 \\
GLM-4.6             & 23 & 19  & 54.8          & 32.5          & \textbf{68.00} \\
\rowcolor{lightgray}
DeepSeek-V3.2       & 10 & 33  & 23.3          & 22.5          & 44.88 \\

\bottomrule
\end{tabular}
\end{table}


\paragraph{Analysis.}
Several findings emerge from comparing automated and human evaluations. First, the model ranking is broadly consistent between the repaied automated metric (All Three) and human judgment, with Gemini-3-Pro leading in both settings and DeepSeek-V3.2 trailing substantially. This suggests that the repaired benchmark's automated metrics are reasonably well-calibrated after \textsc{EigenData}-guided corrections.

Second, comparing the repaired All Three metric against the original BFCL leaderboard scores reveals striking rank inversions that expose the severity of benchmark artifacts prior to repair. Most notably, GPT-5.2 scores only 28.12\% on the original BFCL but achieves 65.0\% human pass rate---a 36.9 percentage-point gap indicating that pre-repair benchmark errors were disproportionately penalizing this model. Conversely, GLM-4.6 attains the highest original BFCL score (68.00\%) yet ranks fourth in human evaluation (54.8\%), suggesting it benefited from benchmark-specific patterns that do not reflect genuine capability. Crucially, the repaired All Three metric corrects these distortions: its model ranking (Gemini-3-Pro $>$ Claude-4.5-Sonnet $>$ GPT-5.2 $\geq$ GLM-4.6 $>$ DeepSeek-V3.2) aligns with the human-evaluation ranking, whereas the original leaderboard ordering does not. This demonstrates that \textsc{EigenData}'s repair pipeline removes systematic biases that previously inflated or deflated model scores, yielding automated metrics whose relative ordering faithfully reflects human judgments of functional correctness.

Third, the sub-metric decomposition in Table~\ref{tab:auto-eval} reveals that multi-turn function calling involves multiple independent failure modes. The average pass rate drops from 57.0\% (Key Function) and 53.8\% (Config Match) individually to 41.0\% when both are required (Config$+$Func), and further to 33.6\% under the full All Three criterion. This 23.4 percentage-point gap between the easiest individual metric and the conjunction indicates that models frequently satisfy one evaluation axis while failing another: a model may invoke the correct functions but with malformed configurations, or produce well-formed calls that miss key functions in the workflow. Notably, the LLM-as-Judge metric (49.0\% average) falls between the individual and conjunction criteria, suggesting that semantic-level assessment captures a complementary signal---detecting cases where both the configuration and function selection are individually plausible but the overall trajectory is logically incoherent. This multi-dimensional failure profile underscores why single-axis evaluation can mask genuine capability gaps, and validates the design of the All Three metric as a stringent but informative composite.

Taken together, these results demonstrate that \textsc{EigenData}'s end-to-end auditing and repair pipeline---coordinated by \textsc{EigenCore} across the \textsc{DatabaseAgent}, \textsc{CodingAgent}, and \textsc{DataAgent}---produces benchmarks that are both more reliable (fewer false negatives from schema mismatches and implementation bugs) and more discriminative (better correlation with human judgments of functional correctness). 





\section{Conclusion}
\label{sec:conclusion}

We have presented EigenData, an integrated, self-evolving platform for function-calling data that coordinates three specialized sub-systems---DatabaseAgent, CodingAgent, and DataAgent---through a top-level orchestrator (EigenCore) to cover the full lifecycle from environment construction to trajectory generation and quality assurance. The platform's multi-agent architecture enables dynamic workflow planning, iterative self-correction, and cross-component feedback, distinguishing it from existing linear data-generation pipelines.

We have demonstrated EigenData's capabilities through a case study on BFCL-V3, where our automated audit identified systematic errors affecting 71.5\% of multi-turn entries across function schemas, implementations, and reference trajectories. The repair pipeline, which routes each error category to the appropriate sub-system with cross-component consistency enforcement, produced a corrected benchmark variant that resolves the rank inversions present in the original leaderboard---most notably, models that were under- or over-estimated are now correctly ranked. On the repaired benchmark, the automated All Three metric aligns with human evaluation in model ranking. These findings validate EigenData's end-to-end auditing approach and motivate the outcome-aware, state-based evaluation protocol enabled by EigenData's VerificationFunctionAgent.

\paragraph{Limitations.}
Several limitations of the current work should be noted. First, our BFCL case study focuses on 200 multi-turn cases; scaling the repair and evaluation to the full benchmark remains future work. Second, while we introduce outcome-aware evaluation through synthesized verification functions, a systematic comparison with other evaluation paradigms (e.g., reward-model-based or simulation-based assessment) is deferred to future work. Third, the current system relies on frontier LLMs for all agent roles; an analysis of cost--quality tradeoffs across model tiers would strengthen practical deployment guidance.


\bibliographystyle{unsrtnat}
\bibliography{references}


\newpage
\appendix

\section{BFCL Error Examples}
\label{sec:bfcl-error-examples}

This appendix provides additional concrete examples of errors identified in the BFCL multi-turn base evaluation suite (Section~4.1). Examples are organized by error category and each is presented in a self-contained box showing the relevant schema, code, ground truth, and explanation.


\begin{figure}[htbp]
\centering
\begin{tcolorbox}[
  colback=red!3,
  colframe=red!60!black,
  fonttitle=\bfseries\small,
  title={Schema Error: Speed unit \texttt{"m/h"} contradicts response unit \texttt{"km/h"} (\texttt{setCruiseControl})},
  boxrule=0.6pt, arc=2pt,
  left=5pt, right=5pt, top=4pt, bottom=4pt
]
\small
\textbf{Entry:} \texttt{multi\_turn\_base\_79}, Turn 2 \hfill \textbf{Domain:} Vehicle Control API

\medskip
\textbf{User prompt:} \textit{``Could you adjust the cruise control to 65 mph\ldots''}

\medskip
\textbf{Schema definition} (\texttt{vehicle\_control.json}):
\begin{lstlisting}[style=jsonstyle, aboveskip=4pt, belowskip=4pt]
{ "name": "setCruiseControl",
  "parameters": {
    "speed": {"type": "float",
              "description": "The speed to set in m/h."}
  },
  "response": {
    "currentSpeed": {"type": "float",
                     "description": "Current speed in km/h."}
}}
\end{lstlisting}

\textbf{Ground truth call:}
\begin{lstlisting}[style=jsonstyle, aboveskip=4pt, belowskip=4pt]
setCruiseControl(speed=65, activate=True, distanceToNextVehicle=100)
\end{lstlisting}

\textbf{Problem:} The input parameter description uses \texttt{"m/h"} (ambiguous---miles/hour or meters/hour?), while the response describes speed in \texttt{"km/h"}. The user says ``65 mph'', but the schema offers two contradictory unit systems. If \texttt{"m/h"} means miles per hour but the output is in km/h, there is an implicit unit conversion that is never documented. If \texttt{"m/h"} means km/h, the abbreviation is wrong.
\end{tcolorbox}
\caption{Schema error: contradictory unit specifications between input parameter (\texttt{"m/h"}) and response (\texttt{"km/h"}).}
\label{fig:bfcl-schema-unit}
\end{figure}

\begin{figure}[htbp]
\centering
\begin{tcolorbox}[
  colback=red!3,
  colframe=red!60!black,
  fonttitle=\bfseries\small,
  title={Schema Error: Documented response key \texttt{"order\_history"} vs.\ actual key \texttt{"history"}},
  boxrule=0.6pt, arc=2pt,
  left=5pt, right=5pt, top=4pt, bottom=4pt
]
\small
\textbf{Function:} \texttt{get\_order\_history} \hfill \textbf{Domain:} Trading Bot API

\medskip
\textbf{Schema} (\texttt{trading\_bot.json}):
\begin{lstlisting}[style=jsonstyle, aboveskip=4pt, belowskip=4pt]
{ "name": "get_order_history",
  "response": {
    "properties": {
      "order_history": {"type": "array",
        "description": "List of orders ID in the order history."}
}}}
\end{lstlisting}

\textbf{Implementation} (\texttt{trading\_bot.py:554}):
\begin{lstlisting}[style=jsonstyle, aboveskip=4pt, belowskip=4pt]
return {"history": list(self.orders.keys())}
\end{lstlisting}

\textbf{Problem:} The schema documents the response key as \texttt{"order\_history"}, but the implementation returns \texttt{"history"}. Any model or downstream code that parses the response using the schema-documented key would fail to find the data.
\end{tcolorbox}
\caption{Schema error: the documented response key \texttt{"order\_history"} does not match the actual implementation key \texttt{"history"}.}
\label{fig:bfcl-schema-response}
\end{figure}


\begin{figure}[htbp]
\centering
\begin{tcolorbox}[
  colback=orange!3,
  colframe=orange!70!black,
  fonttitle=\bfseries\small,
  title={Implementation Error: \texttt{tail(lines=0)} returns ALL lines instead of empty},
  boxrule=0.6pt, arc=2pt,
  left=5pt, right=5pt, top=4pt, bottom=4pt
]
\small
\textbf{File:} \texttt{gorilla\_file\_system.py:563--585}

\medskip
\textbf{Implementation:}
\begin{lstlisting}[style=jsonstyle, aboveskip=4pt, belowskip=4pt]
def tail(self, file_name: str, lines: int = 10):
    content = file._read().splitlines()
    if lines > len(content):
        lines = len(content)
    last_lines = content[-lines:]  # content[-0:] == content[0:]!
    return {"last_lines": "\n".join(last_lines)}
\end{lstlisting}

\textbf{Problem:} In Python, \texttt{content[-0:]} is equivalent to \texttt{content[0:]}, which returns the \emph{entire} list. So \texttt{tail(file, lines=0)} returns \textbf{all} lines instead of no lines. This is a classic off-by-one / boundary condition error. A model calling \texttt{tail} with \texttt{lines=0} would receive unexpected results.
\end{tcolorbox}
\caption{Implementation error: \texttt{tail(lines=0)} returns the entire file due to Python's \texttt{[-0:]} slice semantics.}
\label{fig:bfcl-impl-tail}
\end{figure}

\begin{figure}[htbp]
\centering
\begin{tcolorbox}[
  colback=orange!3,
  colframe=orange!70!black,
  fonttitle=\bfseries\small,
  title={Implementation Error: \texttt{get\_zipcode\_based\_on\_city} --- silent failure on unknown cities},
  boxrule=0.6pt, arc=2pt,
  left=5pt, right=5pt, top=4pt, bottom=4pt
]
\small
\textbf{File:} \texttt{vehicle\_control.py:618--649} \hfill \textbf{Affected entries:} 38 calls across 20 entries

\medskip
\textbf{Implementation:}
\begin{lstlisting}[style=jsonstyle, aboveskip=4pt, belowskip=4pt]
def get_zipcode_based_on_city(self, city: str):
    if city == "Rivermist":
        return {"zipcode": "83214"}
    elif city == "San Francisco":
        return {"zipcode": "94016"}
    # ... more elif chains ...
    else:
        return {"zipcode": "00000"}  # Silent failure!
\end{lstlisting}

\textbf{Problem:} (1)~Case-sensitive exact matching means \texttt{"san francisco"} or \texttt{"SAN FRANCISCO"} returns \texttt{"00000"}. (2)~Unknown cities silently return the invalid zipcode \texttt{"00000"} instead of an error, which then propagates to \texttt{estimate\_distance} producing incorrect distance calculations. The function should either normalize case or return an error for unknown cities.
\end{tcolorbox}
\caption{Implementation error: case-sensitive matching and silent failure with an invalid default zipcode \texttt{"00000"}.}
\label{fig:bfcl-impl-zipcode}
\end{figure}

\begin{figure}[htbp]
\centering
\begin{tcolorbox}[
  colback=orange!3,
  colframe=orange!70!black,
  fonttitle=\bfseries\small,
  title={Implementation Error: \texttt{send\_message} stores messages without sender},
  boxrule=0.6pt, arc=2pt,
  left=5pt, right=5pt, top=4pt, bottom=4pt
]
\small
\textbf{File:} \texttt{message\_api.py:187--196} \hfill \textbf{Affected entries:} 28 entries

\medskip
\textbf{Implementation:}
\begin{lstlisting}[style=jsonstyle, aboveskip=4pt, belowskip=4pt]
def send_message(self, receiver_id: str, message: str):
    message_id = self._generate_id()  # Returns {"new_id": int}!
    self.inbox.append({receiver_id: message})  # No sender field
    self.message_count += 1
    return {"sent_status": True, "message_id": message_id}
\end{lstlisting}

\textbf{\texttt{\_generate\_id}} (\texttt{message\_api.py:104}):
\begin{lstlisting}[style=jsonstyle, aboveskip=4pt, belowskip=4pt]
def _generate_id(self):
    return {"new_id": new_id}  # Returns a dict, not an int
\end{lstlisting}

\textbf{Problem:} Two compounding bugs: (1)~Messages are stored as \texttt{\{receiver\_id: message\}} with no sender field, so \texttt{delete\_message} cannot distinguish between messages from different senders to the same receiver, and \texttt{view\_messages\_sent} returns \emph{all} messages regardless of sender. (2)~\texttt{\_generate\_id} returns a dict \texttt{\{"new\_id": 12345\}} instead of a plain integer, producing a nested dict in the response.
\end{tcolorbox}
\caption{Implementation error: \texttt{send\_message} omits the sender field and \texttt{\_generate\_id} returns a dict instead of an integer.}
\label{fig:bfcl-impl-sendmsg}
\end{figure}


\begin{figure}[htbp]
\centering
\begin{tcolorbox}[
  colback=blue!3,
  colframe=blue!60!black,
  fonttitle=\bfseries\small,
  title={Trajectory Error: Move-then-rename vs.\ rename-then-move (\texttt{multi\_turn\_base\_10})},
  boxrule=0.6pt, arc=2pt,
  left=5pt, right=5pt, top=4pt, bottom=4pt
]
\small
\textbf{Entry:} \texttt{multi\_turn\_base\_10}, Turn 1 \hfill \textbf{Domain:} File System API

\medskip
\textbf{User prompt:} \textit{``Let's move over the project's proposal document into this `Projects' folder, but we'll go ahead and rename it to `final\_proposal\_2024'.''}

\medskip
\textbf{Ground truth} (move first, then rename):
\begin{lstlisting}[style=jsonstyle, aboveskip=4pt, belowskip=4pt]
mv(source='proposal.docx', destination='Projects')
cd(folder='Projects')
mv(source='proposal.docx', destination='final_proposal_2024')
\end{lstlisting}

\textbf{Equally valid alternative} (rename first, then move):
\begin{lstlisting}[style=jsonstyle, aboveskip=4pt, belowskip=4pt]
mv(source='proposal.docx', destination='final_proposal_2024')
mv(source='final_proposal_2024', destination='Projects')
\end{lstlisting}

\textbf{Problem:} Both approaches produce identical filesystem state: \texttt{Projects/final\_proposal\_2024} with the correct content. However, the evaluation requires the ground truth's execution result strings to appear in the model's output. The alternative produces different intermediate strings, causing a false failure. The same pattern affects \texttt{multi\_turn\_base\_24}.
\end{tcolorbox}
\caption{Trajectory error: overly rigid evaluation penalizes a valid rename-then-move strategy that produces identical final state.}
\label{fig:bfcl-traj-ordering}
\end{figure}

\begin{figure}[htbp]
\centering
\begin{tcolorbox}[
  colback=blue!3,
  colframe=blue!60!black,
  fonttitle=\bfseries\small,
  title={Trajectory Error: Float vs.\ integer formatting in tweet (\texttt{multi\_turn\_base\_75})},
  boxrule=0.6pt, arc=2pt,
  left=5pt, right=5pt, top=4pt, bottom=4pt
]
\small
\textbf{Entry:} \texttt{multi\_turn\_base\_75}, Turn 1 \hfill \textbf{Domain:} Vehicle Control + Twitter API

\medskip
\textbf{User prompt:} \textit{``Share a quick update about the tire pressures on Twitter, using the format `Front Left Tire: XXX PSI, \ldots'\,''}

\medskip
\textbf{\texttt{check\_tire\_pressure()} returns} (float values):
\begin{lstlisting}[style=jsonstyle, aboveskip=4pt, belowskip=4pt]
{"frontLeftTirePressure": 32.0, "frontRightTirePressure": 32.0,
 "rearLeftTirePressure": 30.0,  "rearRightTirePressure": 30.0}
\end{lstlisting}

\textbf{Ground truth tweet} (integer formatting):
\begin{lstlisting}[style=jsonstyle, aboveskip=4pt, belowskip=4pt]
post_tweet(content='Front Left Tire: 32 PSI, Front Right Tire:
  32 PSI, Rear Left Tire: 30 PSI, Rear Right Tire: 30 PSI')
\end{lstlisting}

\textbf{Problem:} The API returns float values (\texttt{32.0}, \texttt{30.0}), but the ground truth uses integer formatting (\texttt{32}, \texttt{30}). A model faithfully using the API return values would construct \texttt{"Front Left Tire: 32.0 PSI, \ldots"}. The state checker compares tweet content exactly, so \texttt{32.0} vs \texttt{32} causes a mismatch despite conveying identical information.
\end{tcolorbox}
\caption{Trajectory error: exact string matching penalizes float formatting (\texttt{32.0}) when the ground truth uses integers (\texttt{32}).}
\label{fig:bfcl-traj-float}
\end{figure}

\begin{figure}[htbp]
\centering
\begin{tcolorbox}[
  colback=blue!3,
  colframe=blue!60!black,
  fonttitle=\bfseries\small,
  title={Trajectory Error: \texttt{du()} format mismatch --- no valid solution exists (\texttt{multi\_turn\_base\_28})},
  boxrule=0.6pt, arc=2pt,
  left=5pt, right=5pt, top=4pt, bottom=4pt
]
\small
\textbf{Entry:} \texttt{multi\_turn\_base\_28}, Turn 2 \hfill \textbf{Domain:} File System API

\medskip
\textbf{User prompt:} \textit{``Log the storage usage of the current directory to usage.txt. The content should be the number followed by the word bytes and nothing else.''}

\medskip
\textbf{Ground truth:}
\begin{lstlisting}[style=jsonstyle, aboveskip=4pt, belowskip=4pt]
du(human_readable=True)          # Returns "205.00 B"
touch(file_name='usage.txt')
echo(content='205 bytes', file_name='usage.txt')  # Writes "205 bytes"
\end{lstlisting}

\textbf{\texttt{du()} implementation} (\texttt{gorilla\_file\_system.py:550--561}):
\begin{lstlisting}[style=jsonstyle, aboveskip=4pt, belowskip=4pt]
if human_readable:
    size_str = f"{total_size:.2f} {unit}"  # -> "205.00 B"
else:
    size_str = f"{total_size} bytes"       # -> "205 bytes"
\end{lstlisting}

\textbf{Problem:} The ground truth calls \texttt{du(human\_readable=True)} which returns \texttt{"205.00 B"}, but then writes \texttt{"205 bytes"} (the non-human-readable format). A model using \texttt{du(human\_readable=True)} and faithfully writing its output would write \texttt{"205.00 B"} (state mismatch). A model using \texttt{du()} without the flag gets \texttt{"205 bytes"} (correct file content), but fails the response check because the ground truth expects the \texttt{human\_readable=True} return value. \textbf{No solution can satisfy both checks simultaneously.}
\end{tcolorbox}
\caption{Trajectory error: the ground truth calls \texttt{du(human\_readable=True)} but writes the non-human-readable output, making it impossible to pass both state and response checks.}
\label{fig:bfcl-traj-du}
\end{figure}

\end{document}